\documentclass[12pt]{article}
\usepackage{hyperref}
\usepackage{amsmath, amsfonts, amssymb, mathrsfs}
\usepackage{dcolumn}
\usepackage{caption}
\usepackage{subcaption}
\usepackage{filemod}
\usepackage{natbib}
\usepackage[ruled, vlined]{algorithm2e}
\usepackage{floatrow}
\usepackage{setspace}
\usepackage{verbatim}
\usepackage{graphicx}
\usepackage{booktabs, multicol}
\usepackage[font=small,labelfont=bf]{caption}
\newfloatcommand{capbtabbox}{table}[][\FBwidth]
\usepackage{blindtext}
\usepackage{enumerate}

\hypersetup{
    colorlinks=true,
    linkcolor=blue, 
    urlcolor=black,
    citecolor=blue, 
}

\title{\vspace{-1cm} Vecchia approximated\\Bayesian heteroskedastic Gaussian processes}
\author{Parul V. Patil \thanks{Corresponding author: Department of Statistics, 
        Virginia Tech, {\tt parulvijay@vt.edu}} 
    \and Robert B. Gramacy \thanks{Department of Statistics, Virginia Tech}
    \and Cayelan C. Carey \thanks{Department of Biological Sciences and Center for Ecosystem Forecasting, Virginia Tech}
    \and R. Quinn Thomas \thanks{Departments of Forest Resources \& Environmental Conservation and Biological Sciences and Center for Ecosystem Forecasting, Virginia Tech}}
\date{\today}

\usepackage{hyperref}
\begin{document}

 \newcommand{\KN}{K_{\theta_Y}(X_N)}
  \newcommand{\Kn}{K_{\theta_Y}(X_n)}
  
 \newcommand{\KNL}{K_{\theta_\lambda}(X_N)}
 \newcommand{\Knl}{K_{\theta_\lambda}(X_n)}

\newcommand{\tl}{\theta_{\lambda}}
\newcommand{\taul}{\hat{\tau}_{\lambda}^2}
\newcommand{\gl}{g_{\lambda}}

\newcommand{\ty}{\theta_Y}
\newcommand{\tauy}{\hat{\tau}^2_N}
 
\newcommand{\Ux}{U}
\newcommand{\Uxt}{U^\top}
\newcommand{\Uxinv}{ (U_x U_x^\top)^{-1}}

\newcommand{\Uxn}{U_n}
\newcommand{\Uxtn}{U_n^\top}

\newcommand{\Uxl}{U_n^{(\lambda)}}
\newcommand{\Uxlt}{U_n^{(\lambda)}\top}

\newcommand{\Uxninv}{ (U_n U_n^\top)^{-1}}
\newcommand{\UNinv}{ (U_N U_N^\top)^{-1}}
\newcommand{\Uxlinv}{( U_n^{(\lambda)} U_n^{(\lambda)\top})^{-1}}

\newcommand{\Uell}{U_\mathcal{X}^{(\lambda)}}
\newcommand{\Uellt}{U_\mathcal{X}^{(\lambda)^\top}}
\newcommand{\Uxell}{U_{n, \mathcal{X}}^{(\lambda)\top}}

\newcommand{\bln}{\mathbf{\lambda_n}}

\vspace{-0.5cm}
\maketitle

\vspace{-1.5cm}
\begin{abstract}    

Many computer simulations are stochastic and exhibit input dependent noise. In
such situations, heteroskedastic Gaussian processes ({\tt hetGP}s) make ideal
surrogates as they estimate a latent, non-constant variance. However, existing
{\tt hetGP} implementations are unable to deal with large simulation campaigns
and use point-estimates for all unknown quantities, including latent
variances. This limits applicability to small experiments and undercuts
uncertainty.  We propose a Bayesian {\tt hetGP} using elliptical slice
sampling (ESS) for posterior variance integration, and the Vecchia
approximation to circumvent computational bottlenecks. We show good
performance for our upgraded {\tt hetGP} capability, compared to alternatives,
on a benchmark example and a motivating corpus of more than 9-million lake
temperature simulations. An open source implementation is provided as {\tt
bhetGP} on CRAN.

\end{abstract}


\section{Introduction}
\label{sec:intro}

Computer simulation experiments are common in scientific fields where
physical/field experimentation may be challenging
\citep[e.g.,][]{booker1998design,santner2018design}. Examples include ecology
\citep{johnson2008microcolony}, epidemiology \citep{hu2017sequential,
fadikar2018calibrating}, aeronautics \citep{mehta2014modeling} and engineering
\citep{zhang2015microstructure}. It used to be that computer codes simulating
complex phenomena were deterministic, meaning identical inputs yield
{the} same output.  Often they were solvers of systems of
differential equations. Modern simulations are increasingly stochastic,
utilizing pseudo-random number generators in some aspect, either via Monte
Carlo (MC) for numerical quadrature
\citep[e.g.,][]{mehta2014modeling,herbei2014estimating}, or to virtualize a
system which is inherently stochastic
\citep[e.g.,][]{xie2012assemble,fadikar2018calibrating}.
\citet{baker2022analyzing} provide a recent review of this setting,
highlighting opportunities and challenges.

Our interest in stochastic simulation lies in an application of the General
Lake Model \citep[GLM;][]{gmd-12-473-2019}\footnote{We apologize for the
acronym; there is no connection to generalized linear modeling.} to reservoir
temperature forecasting. The open-source GLM model has been applied to
thousands of lakes around the world to simulate water quality
\citep{read2014simulating, bruce2018multi}. Water temperature impacts the
formation of phytoplankton blooms which can adversely impact drinking water
quality \citep{carey2012eco}.  GLM simulations are technically deterministic,
but when driven by data products that summarize uncertainties in
weather/climate variables, they propagate those stochastic effects.  Here, we
follow \citet{holthuijzen2024synthesizing} and study GLM as driven by an
ensemble of forecasts generated by the National Oceanic and Atmospheric
Administration (NOAA) Global Ensemble Forecast System
\citep[NOAAGEFS;][]{hamill2022reanalysis}\footnote{See
\url{https://www.ncei.noaa.gov/products/weather-climate-models/global-ensemble-forecast}.}.
Going forward, we refer to NOAA-GLM.
\begin{figure}[ht!]
\centering
\includegraphics[scale = 0.52, trim=3 10 28 50, clip =TRUE]{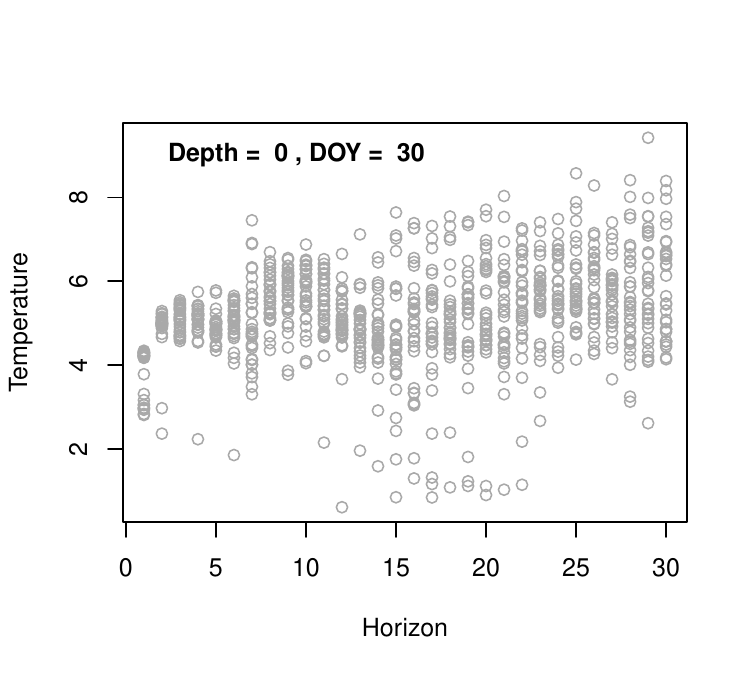}
\includegraphics[scale = 0.52, trim=30 10 28 50, clip =TRUE]{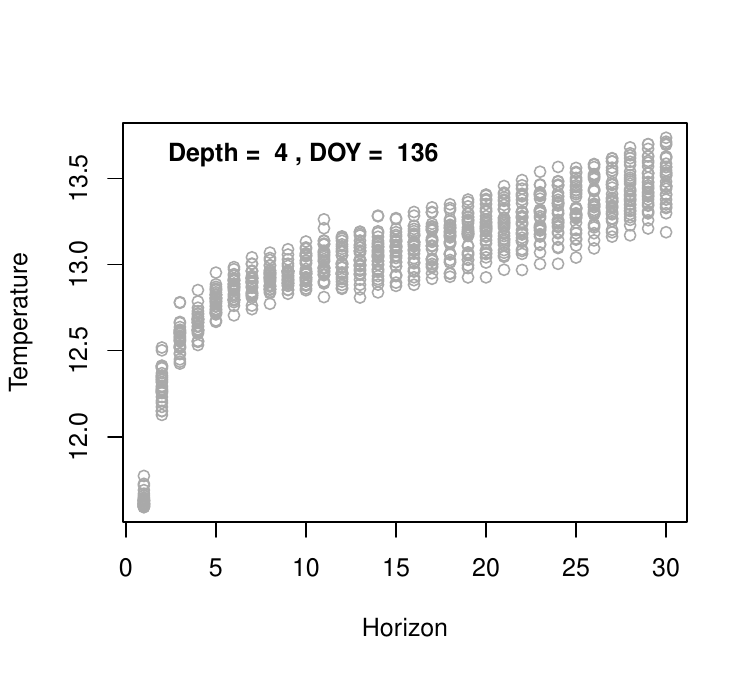}
\includegraphics[scale = 0.52, trim=30 10 30 50, clip =TRUE]{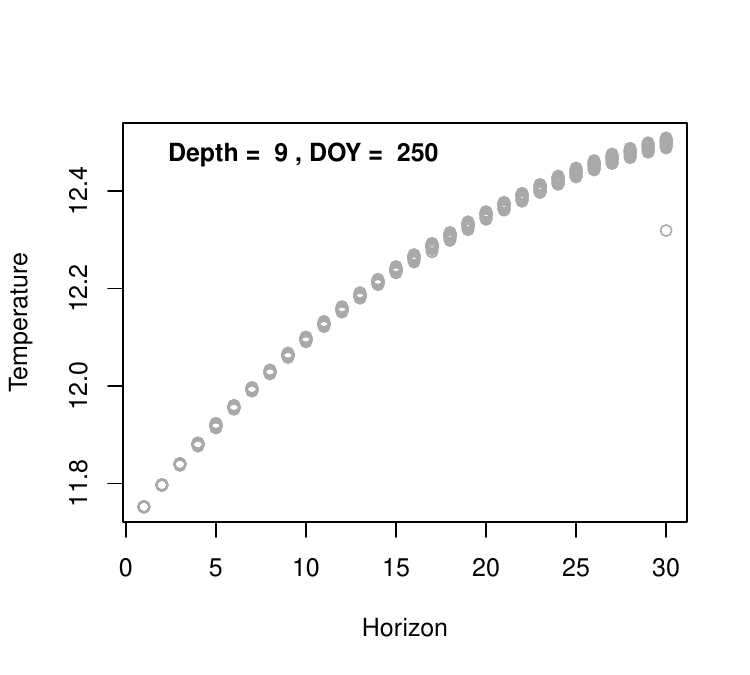}
\caption{NOAA-GLM simulations ahead to a horizon of 30 days, varying day of
year (DOY) in 2022 and depth from the surface
of the lake in meters.}
\label{fig:lakes_intro}
\end{figure}
Example NOAA-GLM simulated temperatures ($y$-axes) are provided in
Figure~\ref{fig:lakes_intro}; more detail is provided in Section
\ref{sec:glm}. For now, simply notice the non-linear relationship, varying
noise level over inputs, and large number of outputs corresponding to
temperature forecasts at {three days of the year (DOY) over
thirty forecasting horizon days}.  Across the three panels there are 2730 gray
circles. \citet{holthuijzen2024synthesizing} provide us with three years worth
simulations for ten depths, {i.e., $< 10{,}000\times$ more gray
circles.}

It can be advantageous to abstract a corpus of simulations, like those in
Figure~\ref{fig:lakes_intro}, with a statistical model relating inputs to
outputs.  Such abstractions are known as {\em surrogates} or {\em emulators}
in the computer modeling literature
\citep{santner2018design,gramacy2020surrogates}. One reason to fit a surrogate
is to economize on expensive simulations, keeping experiment sizes down, or to
cope with limited access to simulation resources. A good surrogate can furnish
predictions with appropriate uncertainty quantification (UQ) in lieu of
additional simulations, tell you where to put new runs in order to accomplish
some design goal like identifying safe/efficient operating conditions
\citep[e.g.,][]{booth2025contour}, or help calibrate simulations to field data
\citep[e.g.,][]{kennedy2001bayesian}.

Although many statistical models make good surrogates -- any regression model
would be appropriate -- Gaussian processes \citep[GPs;
e.g.,][]{williams2006gaussian} are canonical because they are non-parametric,
nonlinear and provide accurate and smooth predictions {(using
typical kernels based on inverse Euclidean distance)} along with
well-calibrated UQ.  Yet for some situations GPs lack the flexibility required
to capture the salient dynamics being simulated. Input-dependent noise, like
in Figure \ref{fig:lakes_intro}, is one such setting. As
\citet{baker2022analyzing} explains, stochastic processes are often
heteroskedastic (exhibit input-dependent noise) not homoskedastic (constant
noise for all inputs). Another downside to GPs involves computation. With $N$
training-data input-output pairs, GP modeling involves decomposing a dense
$N\times N$ covariance matrix, incurring costs in $\mathcal{O}(N^3)$ flops.
This limits $N$ to the small thousands. Noisy simulations exacerbate this
bottleneck because bigger $N$ is needed to identify the signal.

\citet{goldberg:williams:bishop:1998} were the first to propose a
heteroskedastic GP, coupling two GPs: one for means and another for variances.
But their Bayesian {Markov chain Monte Carlo (MCMC)} based
inferential toolkit, involving independent Metropolis proposals for $N$ latent
variance variables, was cumbersome, limiting $N$ to the small hundreds. About a
decade later, several thriftier variants arrived. \citet{binois2018practical}
provide a review toward appropriating two key ingredients from that line of
research for their own method: 1) reduced sufficient statistics from training
designs that deploy replication ($N$ simulations at $n \ll N$ unique inputs),
first proposed as {\em stochastic kriging}
\citep[SK;][]{ankennman:nelson:staum:2010}; 2) point-estimation of latent
variances rather than full posterior sampling, e.g., via expectation
maximization \citep[EM;][]{kersting:etal:2007}.
\citeauthor{binois2018practical}~glued the two ideas together with a
well-known linear algebra trick known as the Woodbury identity, requiring only
$\mathcal{O}(n^3)$ flops.  Their {\tt hetGP} is the only method with an
implementation in public software for {\sf R} \citep{binois2021hetgp} and {\sf
Python} \citep{o2025hetgpy}.

However, there are two downsides to {\tt hetGP}, both broadly and for
\citet{holthuijzen2024synthesizing}'s NOAA-GLM simulations.  One is that
point-inference of unknown quantities, particularly variances, undercuts UQ.
Another is the sheer size of modern stochastic simulation
campaigns.\footnote{{A simulation campaign is the plan, or
design, of the experiment coupled with a strategy for partitioning and
scheduling that work in a (usually shared) high performance computing (HPC)
environment.}} NOAA's thirty-one member ensemble means a 31-fold degree of
replication ($n = N/31$). If Figure \ref{fig:lakes_intro} comprised the entire
campaign, showing $n=90$ unique DOY and depth combinations, then {\tt hetGP}
would work fine. But three years of simulations, at ten depths, and a 30-day
horizon means $n > 300{,}000$. You can't build an $n
\times n$ matrix of that size on workstation, let alone decompose one.

\citet{holthuijzen2024synthesizing} were able to circumvent the second
(computational) downside but at the expense of the first (modeling and
inferential fidelity).  They used the scaled Vecchia approximation
\citep{katzfuss2022scaled} to induce sparsity in the (inverse) covariance
matrices involved in the two (mean and variance) GPs. The downgrade in
fidelity comes from a decoupling of the two GPs in order to offload
computations to a software library.  They had to lean heavily
{on} independent, moment-based inference as
{opposed} to the joint, likelihood-based approach of {\tt
hetGP}.

Here we show that it's possible to perform fully Bayesian posterior
integration for {\tt hetGP} in a way that is both computationally tractable,
and provides more accurate predictions and full UQ.  In addition to the
ingredients already outlined above -- (1) coupled GPs, (2) replication-based
sufficient statistics, and (3) Vecchia approximation -- we provide a key, new
ingredient that {makes} this possible: (4) elliptical slice
sampling \citep[ESS;][]{murray2010elliptical}. ESS is ideal for Bayesian MCMC
under Gaussian priors.  However, to say ``we use ESS'' is a vast
over-simplification. Nobody has ever put these together before, because it's
not {easy}.  Although we draw inspiration from deep GPs
\citep{sauer2023vecchia}, working with replication-based sufficient statistics
in this context presents new challenges.  There are many pitfalls.  E.g.,
using $\mathcal{O}(n)$ sufficient statistics is key to getting it to work.
Vecchia on the original $O(N)$ values, as would be conventional, is too
sparse.

To help clarify the pedigree of ideas involved we have adopted an
unconventional layout for the paper. Each of the following sections introduces
an existing method with review.  Then, we explain the novel way it's been
adapted to suit our large-scale, Bayesian {\tt hetGP} setting.
For example, we review \citeauthor{goldberg:williams:bishop:1998}'s coupled-GP
mean and variance model {in} Section \ref{sec:hetgp}.  Then we
explain how ESS can be used for full posterior inference (which is new).  In
Section \ref{sec:thrifty} we review how \citeauthor{binois2018practical}'s
replicate-based Woodbury likelihood can make inference for latent variances
more efficient both statistically and computationally.  Then we show how to
incorporate that into the fully Bayesian/ESS setup instead of maximizing the
log likelihood (new).  Finally, in Section \ref{sec:vec} we review
\citeauthor{katzfuss2022scaled}'s Vecchia approximation.  Then we explain how
it can inserted into the Woodbury likelihood (new). Along the way we provide
illustrations/visuals to demonstrate how {these} new
ingredients work together, and represent an advance on the state-of-the-art. A
full empirical evaluation is provided in Section \ref{sec:bench}, including
benchmark data and NOAA-GLM lake temperatures.  We conclude with a brief
discussion in Section \ref{sec:discussion}.

\section{Heteroskedastic Gaussian processes}\label{sec:hetgp}

We first introduce GP basics, beginning with a standard, homoskedastic setup
as a prior over means.  Review continues with
\citet{goldberg:williams:bishop:1998}'s heteroskedastic extension via an
additional GP prior over variances.  We then depart from
\citeauthor{goldberg:williams:bishop:1998}~to introduce our first novel
component: deploying ESS for posterior or sampling of latent variances rather
element-wise Metropolis.

\subsection{Review}\label{sec:hetgpR}



Let $f: \mathbb{R}^d \mapsto \mathbb{R}$ abstract a (possibly stochastic)
computer model simulation from $d$ inputs $x$ to one output $y \sim f(x)$. Now
consider a simulation campaign of size $N$.  Let $X_N$ be an $N \times d$
matrix collecting inputs, and $Y_N \sim f(X_N)$ comprise a column vector of
outputs. A GP model, or prior, for these data presumes that they may be
expressed as draw from a multivariate normal distribution (MVN): $Y_N \sim
\mathcal{N}_N(\mu(X_N), \Sigma(X_N))$.  Many models fit this description.
{Simple linear regression (SLR) follows via  $\mu(X_N) = (1, X_N)
\beta$ and $\Sigma(X_N) = \sigma^2 \mathbb{I}_N$. However, it is also possible
to specify a prior over linear functions purely through $\Sigma(X_N)$}. 
When
people think of GPs {in a surrogate modeling context, a
typical starting-off point is} $\mu(X_N) = 0$, and $\Sigma(X_N)$ specified
so that the $Y_N$ are less correlated when their inputs are farther apart.
{There are exceptions, e.g., with means that are linear over
features (like SLR) and covariances that are not distance-based (e.g.,
periodic). We stick to a basic zero-mean/distance-based $\Sigma(X_N)$ setup
here.}

One common (homoskedastic) GP specification \citep[e.g.,][Chapter
5]{gramacy2020surrogates} is as follows.
$$
Y_N \sim \mathcal{N}_N \left( 0 , \tau^2 (K_{\theta} (X_N) + g \mathbb{I}_N) \right) 
\quad \mbox{ where } \quad
K_{\theta} (X_N)^{ij} = k(x_i, x_j) = \exp \left\{- \sum_{k=1}^d \frac{(x_{ik} - x_{jk})^2}{\theta_k} \right\},
$$
introducing scale $\tau^2$, nugget $g$ and lengthscale $\theta=(\theta_1,
\dots, \theta_d)$ hyperparameters.  Many kernels $k(\cdot, \cdot)$ and 
hyperparameters work well \citep[e.g., Matern,][]{stein2012interpolation} and
this choice not directly relevant to the nature of our contribution.  An MVN
density, expressed for $Y_N$ given unknown hyperparameters, defines a log
likelihood -- a special case of Eq.~\eqref{eq:logl} coming momentarily -- that
can be used for learning, either via maximization, MCMC, or otherwise.
{Here, $g$ is used to model the process noise. However, when}
$f$ is deterministic one can take {$g= \varepsilon$}, although
it can sometimes be advantageous estimate {a larger} nugget
{regardless} \citep{gramacy2012cases}. A non-zero setting
specifies an IID noise component with constant level $\tau^2 g$ for all
inputs.

Conditional on hyperparameters, prediction for $\mathcal{Y}(\mathcal{X})$
follows by extending the MVN relationship from training to testing
$\mathcal{X}$, an $n_p \times d$ matrix, by ``stacking'':
\begin{equation}\label{eq:stack}
\begin{bmatrix} 
Y_N \\  
\mathcal{Y} \\ 
\end{bmatrix}
    \sim \mathcal{N}_{N + n_p}
    \left(0, \Sigma_{\text{stack}}  \right)
    \quad \text{where} \quad
    \Sigma_{\text{stack}} = 
     \Sigma \left( 
     \begin{bmatrix} X_N \\ \mathcal{X} \\ \end{bmatrix}
     \right) = 
     \begin{bmatrix} 
         K_{\theta}(X_N) + g \mathbb{I}_n & K_\theta({X_N, \mathcal{X}}) \\ 
             K_\theta({\mathcal{X}, X_N}) & K_\theta(\mathcal{X}, \mathcal{X})\\
        \end{bmatrix}.
        \;
\end{equation}  
Then, standard MVN conditioning yields $\mathcal{Y}(\mathcal{X}) \mid Y_N, X_N
\sim \mathcal{N} (\mu_N(\mathcal{X}), \Sigma_N(\mathcal{X}))$ in closed form
as:
\begin{align}
\mu_N(\mathcal{X}) & 
= {K_{\theta}(\mathcal{X}, X_N) }(K_{\theta} (X_N) + g \mathbb{I}_N)^{-1} Y_N \label{eq:gppred} \\
\Sigma_N(\mathcal{X}) & = \tau^2 \left( K_{\theta}(\mathcal{X}, \mathcal{X}) + \nonumber
g \mathbb{I}_{n_p} - K_{\theta}(\mathcal{X}, X_N) (K_{\theta}(X_N) + g \mathbb{I}_N) ^{-1}
K_{\theta}(X_N, \mathcal{X}) \right).
\end{align}
These are sometimes called the ``kriging equations''
\citep{matheron1963principles}. 
\begin{figure}[ht!]
\centering
\includegraphics[scale = 0.58, trim=2 65 25 20, clip =TRUE]{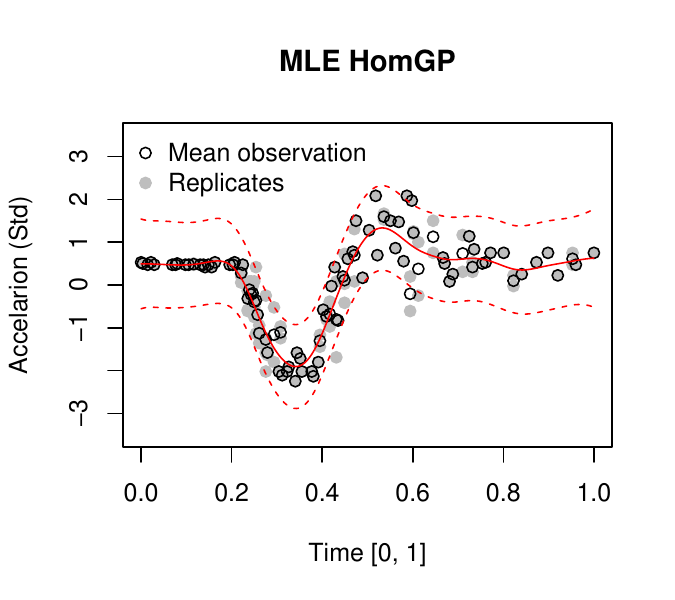}
\includegraphics[scale = 0.58, trim=58 65 25 20, clip =TRUE]{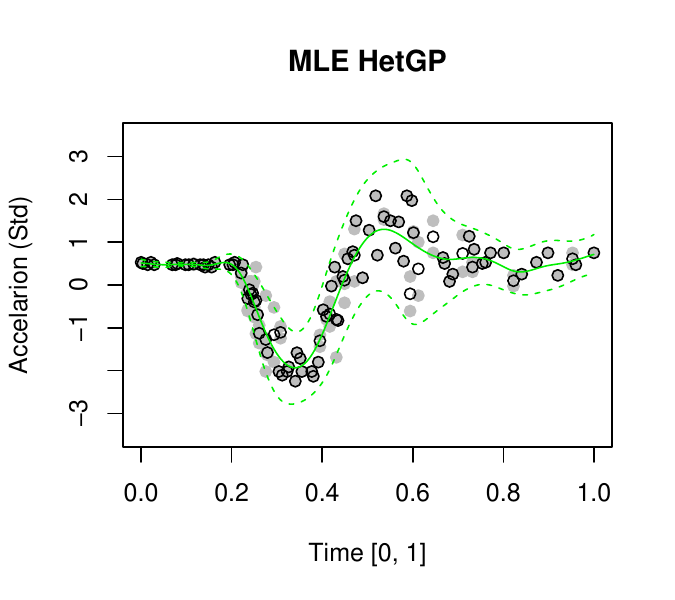}
\includegraphics[scale = 0.58, trim=58 65 30 20, clip =TRUE]{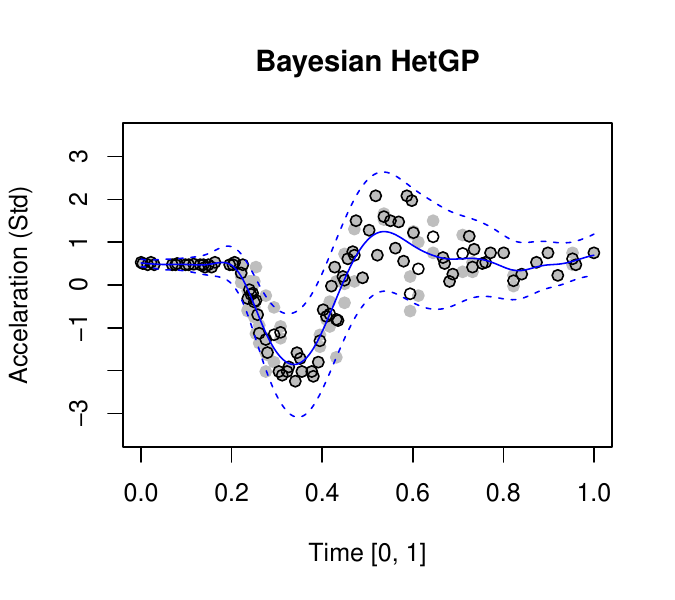}
\includegraphics[scale = 0.58, trim=2 0 25 50, clip =TRUE]{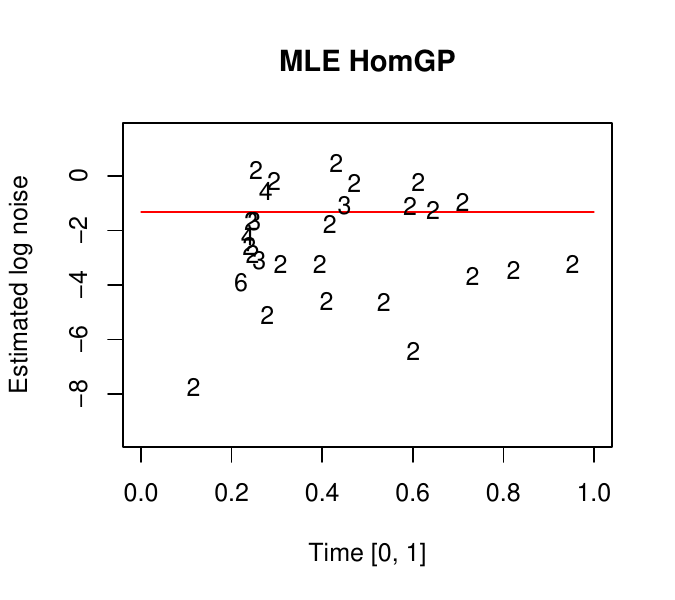}
\includegraphics[scale = 0.58, trim=58 0 25 50, clip =TRUE]{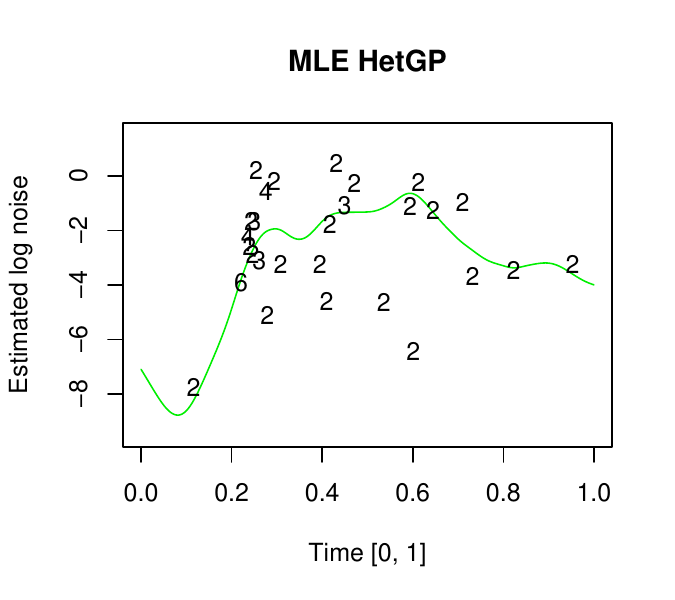}
\includegraphics[scale = 0.58, trim=58 0 30 50, clip =TRUE]{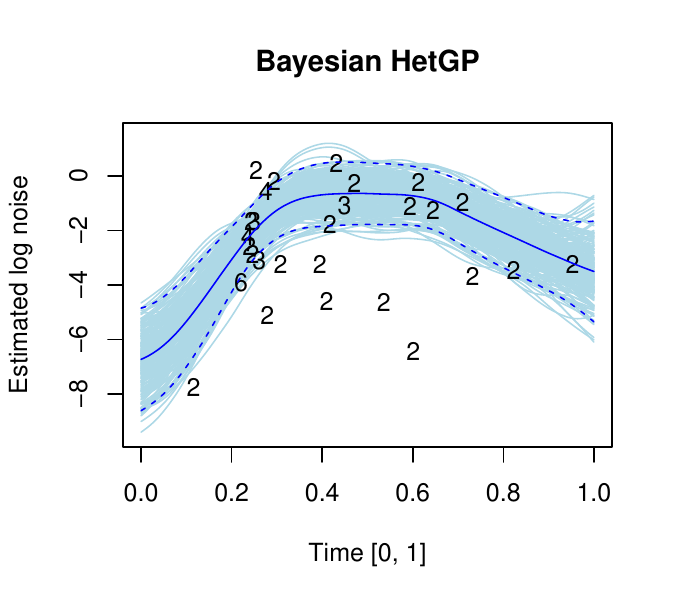}
\caption{{\em Top:} Fits on the motorcycle data via (MLE, homoskedastic) GP,
{\tt hetGP}, and our Bayesian {\tt hetGP}. {\em Bottom:} estimated log-noise,
numbers indicate empirical variances based on that many replicates.}
\label{fig:mcycle}
\end{figure}
The {\em top-left} panel of Figure
\ref{fig:mcycle} provides an illustration on the motorcycle data \citep{mass}.
We shall use this as a running example, returning to other panels over the
next several pages. 
The solid line is $\mu_N(\mathcal{X})$ and the dashed 90\% error-bars come
from $\Phi^{-1}_{0.90}$ and the diagonal of $\Sigma_N(\mathcal{X})$. Maximum
likelihood estimated (MLE) hyperparameters were used.  The {\em bottom-left}
panel plots $\log \tau^2 g$ horizontally.  Each number indicates the
multiplicity of replicates providing a log residual sum-of-squares estimate.

\citet{goldberg:williams:bishop:1998}'s idea for a heteroskedastic GPs
({\tt hetGP} going forward) extends the constant IID component as $g
\mathbb{I}_N \rightarrow \Lambda_N$, where $\Lambda_N$ is a diagonal matrix
storing latent $\lambda_1, \dots, \lambda_N$ variables.  These quantities, or
their logarithms to ensure positivity, are placed under a GP prior
\begin{align}
Y_N & \sim \mathcal{N} \left( 0 , \tau^{2}_N \left( \KN + \Lambda_N \right) \right) \label{eq:hetgp} \\
\log\mathrm{vec}\, \Lambda_N & 
\sim \mathcal{N} \left( 0 , \tau^{2}_\lambda \left( {K_{\theta_\lambda}(X_N)} + g_\lambda \mathbb{I}_N\right) \right), \label{eq:lamp}
\end{align} 
completing a hierarchical specification for nonlinear mean and variance. The
marginal log-likelihood for ($Y_N \mid X_N, \Lambda_N$), used to infer
$\Lambda_N$ along with kernel hyperparameters including
$\tau_\lambda^2$, $\theta_\lambda$ and $g_\lambda$, is
\begin{align}
\log \mathcal{L_N}(Y_N \mid X_N, \Lambda_N) 
&\propto - \frac{N}{2} \log \hat{\tau}^2_N  - \frac{1}{2} \log | \KN + \Lambda_N | \label{eq:logl} \\
\mbox{where } \quad \hat{\tau}^2_N  &= \frac{1}{N+\alpha_Y} (Y_N^\top (K_{\theta_Y} (X_N) + \Lambda_N)^{-1} Y_N  + \beta_Y). \nonumber
\end{align}
The quantity $\hat{\tau}^{2}_N$ arises after integrating out $\tau^2$ under a
$\mathrm{IG}(\alpha_Y/2, \beta_Y/2)$ prior, allowing us to avoid
repeating a similar expression later in Section \ref{sec:hier}.  A reference
prior ($\alpha_Y =
\beta_Y = 0$) corresponds to a profile MLE approach. Taking $\lambda_i = g$
for all $i$ reduces to the homoskedastic setting.
    
\citeauthor{goldberg:williams:bishop:1998}~{described} a
Metropolis-within-Gibbs scheme with independent random-walk proposals for
$\lambda_i$, for $i=1,
\dots, N$. Each evaluation of the likelihood
\eqref{eq:logl} requires $\mathcal{O}(N^3)$ flops for the determinant and
inverse of an $N \times N$ matrix.  Poor mixing meant that $N \approx 100$ was
barely feasible {(see Appendix \ref{app: MH})}.  When predicting under {\tt hetGP}, follow
Eq.~\eqref{eq:gppred} except with $g\mathbb{I}_N
\leftarrow
\Lambda_N$ and $g\mathbb{I}_{n_p} \leftarrow \Lambda(\mathcal{X})$, where this
latter quantity comes from a second application of Eq.~\eqref{eq:gppred} but
for $\log \mathrm{vec}\, \Lambda_N$ values instead of $Y_N$, and with other
$\lambda$-subscripted hyperparameters \eqref{eq:lamp}. The top-middle panel of
Figure \ref{fig:mcycle} provides an example {\tt hetGP} surface, however we
did not use \citeauthor{goldberg:williams:bishop:1998}'s MCMC for this fit. We
maximized the likelihood \eqref{eq:logl}; details {are
provided} in Section \ref{sec:woodbury}. Notice the input-dependent width of
the error-bars, derived from diagonal of $\tau^2\mathbb{E}\{\log
\mathrm{vec}\, \Lambda(\mathcal{X})\}$, shown separately in the bottom panel.

\subsection{Elliptical slice sampling}\label{sec:hetgpN}

Our first novel contribution involves the application of an MCMC technique
that is, in our opinion, underappreciated.  Elliptical slice sampling
\citep[ESS;][]{murray2010elliptical} is designed for high dimensional
posterior sampling under {an} MVN prior.  This is exactly the
situation for $\Lambda_N$ in Eq.~\eqref{eq:hetgp}, but the potential
application for heteroskedastic modeling was not recognized by either the
MCMC/ESS or the surrogate modeling/{\tt hetGP} communities.  Most of the {\tt
hetGP} literature pre-dates ESS, as we review later in Section
\ref{sec:thrifty}, and has all but abandoned MCMC sampling in favor of
point-estimation via optimization. It's time, we think, for an overhaul that
returns {\tt hetGP} to it's Bayesian roots.

Two aspects of ESS combine to make it ideally suited to
Eqs.~(\ref{eq:lamp}--\ref{eq:logl}), and other related latent-GP contexts like
classification \citep{cooper2025modernizing} and deep GPs
\citep{sauer2023active} from which we have drawn inspiration: 1) the Markov
chain is joint for the entirety of $\mathrm{vec}\, \Lambda_N$, so the random
walk is $N$-dimensional; 2) you never reject despite only drawing from an MVN
once, thereby avoiding a ``sticky'' chain.  The essence is as follows.  Let
$\Lambda_N^{(t-1)}$ denote the current sample of latent variances at $X_N$.
{Sample} $\log \mathrm{vec}\,
\Lambda_N^{\mathrm{{prior}}}$ from {its} prior
\eqref{eq:lamp} and draw a random angle $\gamma \sim \mathrm{Unif}(0, 2\pi)$.
Combine {these two} $\Lambda_N$-values by tracing out angle
$\gamma$ of an ellipse passing through them:
\begin{equation}
\label{eq:ess2}
\log \Lambda_N^\star  = \log \Lambda_N^{(t-1)} \cos \gamma + \log  \Lambda_N^{{\mathrm{prior}}} \sin \gamma.
\end{equation}
Then, conduct ordinary slice sampling \citep{neal2003slice} on that ellipse.
Compare $\Lambda_N^\star$ to $\Lambda_N^{(t-1)}$ via a Metropolis-style
likelihood \eqref{eq:logl} ratio
\begin{equation}\label{eq:ess3}
\alpha = \min \left( 1, 
\frac{\mathcal{L}(Y_N \mid X_N, \Lambda_N^\star)}{\mathcal{L}(Y_N \mid X_N, 
\Lambda_N^{(t-1)})} \right).
\end{equation}
If accepted, you have a new $\Lambda_N^{(t)} \leftarrow \Lambda_N^\star$.  If
not, narrow the angle on the ellipse
\begin{equation}\label{eq:ess4}
\gamma \sim \text{Unif} (\gamma_{\min}, \gamma_{\max}) \quad 
\text{where, } \gamma_{\min} = \gamma \text{ if } \gamma < 0 \text{ and } \gamma_{\max} = \gamma 
\text{ otherwise},
\end{equation}
and repeat from Eq.~\eqref{eq:ess2}.  That last part is key.  Whereas Metropolis
would take the previous value as the next one upon rejecting, creating a
sticky chain, ESS adapts and moves $\Lambda_N^\star$ closer to
$\Lambda_N^{(t)}$.  In most applications acceptances come after a small
handful of iterations.

\begin{figure}[ht!]
\centering
\includegraphics[scale = 0.6,trim=0 15 28 15]{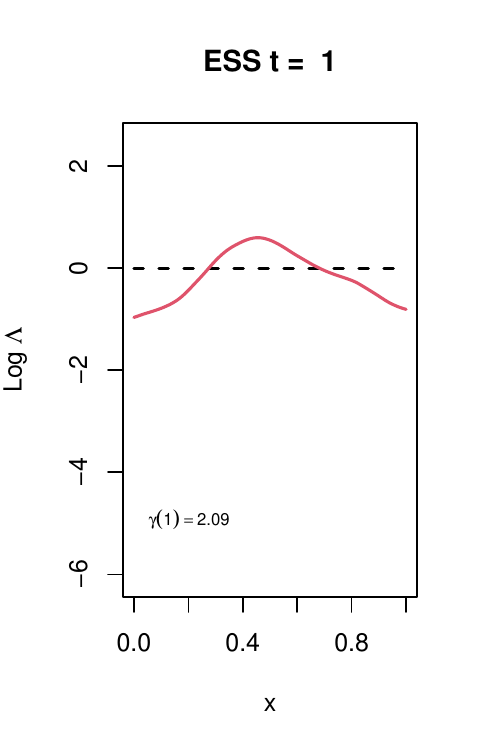}
\includegraphics[scale = 0.6,trim=55 15 28 15,clip=TRUE]{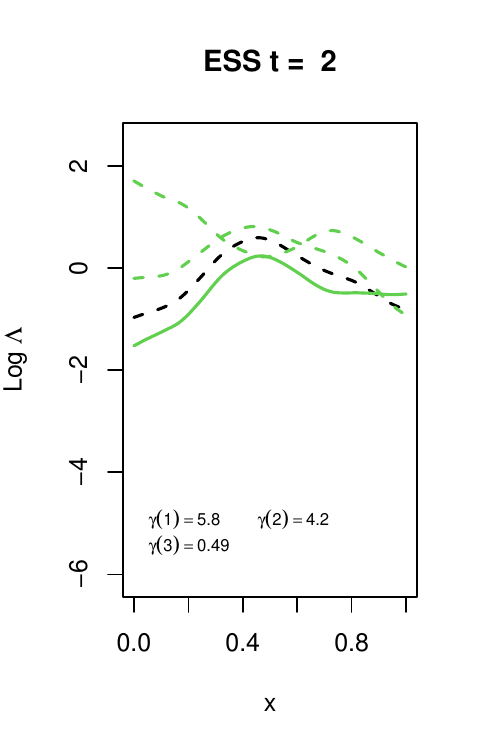}
\includegraphics[scale = 0.6,trim=55 15 28 15,clip=TRUE]{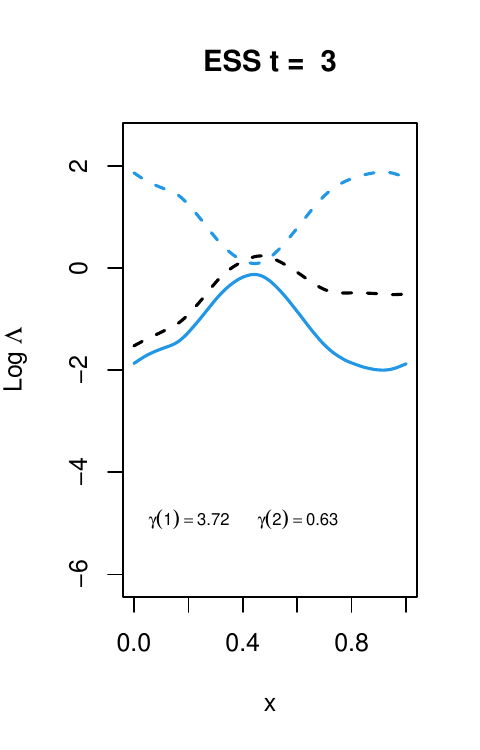}
\includegraphics[scale = 0.6,trim=55 15 28 15,clip=TRUE]{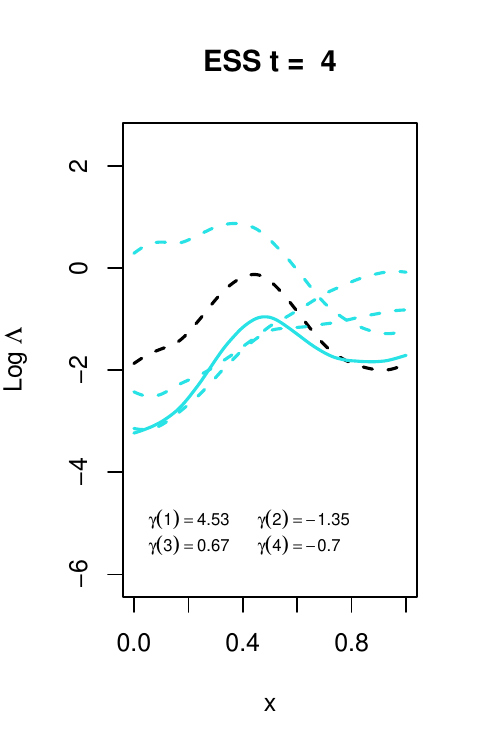}
\includegraphics[scale = 0.6,trim=55 15 28 15,clip=TRUE]{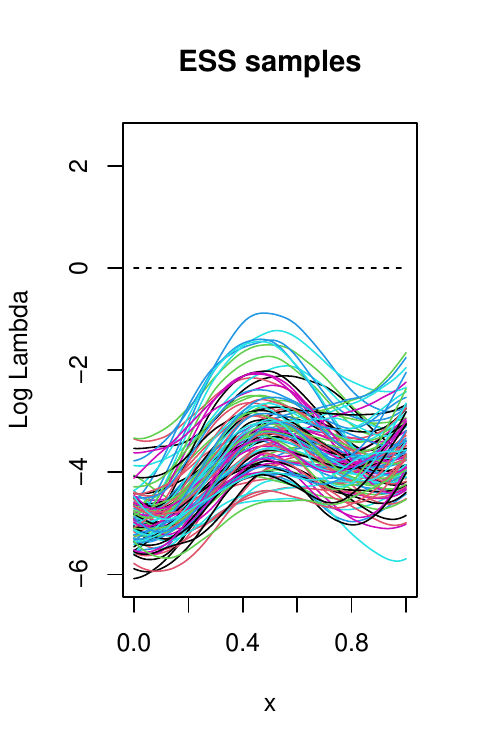}
\caption{ESS iterations {($t$)} in 1D. Black lines indicate the initial value, dashed
lines are rejected proposals and solid lines are the final accepted ones. The
last panel shows all ESS samples after burn-in and thinning.}
\label{fig:ess_s2}
\end{figure}

Figure \ref{fig:ess_s2} showcases an example sequence of ESS draws in its
first four panels. The dashed black line is the previously accepted sample at
iteration $t-1$, with $t=0$ level at $\log \Lambda_N^{(0)} = 0_N$.  Dashed colored
lines represent rejected and refined {\em within}-ESS $\log \Lambda_N^\star$
proposals (\ref{eq:ess2}--\ref{eq:ess4}), until the final accepted
$\log \Lambda_N^{(t)}$ one (solid line).  So at $t=1$ the very
first $\log \Lambda_N^\star$ was accepted, but at $t=2$ there were two rejected
proposals, and subsequent refinements of $\gamma$, before acceptance.

Fast forwarding, the resulting collection of $\log \Lambda_N^{(t)}$ values,
for $t=1,\dots,T$ (after burning in 1000 samples and thinning the chain by
10), shown in the final panel, comprise an empirical distribution that can be
filtered {to any downstream inference, such as prediction}
\eqref{eq:gppred}.  This is plotted for the motorcycle data in the {\em right}
column of Figure \ref{fig:mcycle}. {A} full distribution of
log noises {is shown} on the bottom. Notice also that the mean
of this surface, indicated by the thick solid blue line, is much less
``wiggly'' than the most probable setting to it's left, in the {\em middle}
panel. {Noise estimates are smoothed, leading to larger
inferred variance in some regions (e.g., $x \approx$ 0.35). Although the
posterior incorporates the possibility of smaller noise levels, everywhere,
mostly posterior variance samples are larger than the posterior-maximizing
setting.} Consequently, the predictive surface in the panel above, which
synthesizes all of the samples for all unknowns, looks smoother and has
generally wider error-bars {than in the {\em middle} panel}. 

\section{Thrifty inference under replication}
\label{sec:thrifty}

Since \cite{goldberg:williams:bishop:1998}, several advances brought practical
tractability to the {\tt hetGP} framework. Our review here is not exhaustive,
focusing instead on broad themes, landing on an important identity that we
appropriate for our fully Bayesian, ESS-based scheme.

One line of inquiry came from machine learning (ML), beginning with
\citet{kersting:etal:2007} who replaced posterior integration with expectation
maximization \citep[EM;][]{dempster1977maximum}. The flow involved estimating
empirical noise levels via residuals from a preliminary GP, which resulted
in a non-smooth noise process. \citet{quadrianto:etal:2009} followed with a
penalized likelihood/Maximum-a-posteriori (MAP) approach, whereas
\citet{lazaro-gredilla:tsitas:2011} deployed variational inference.  These led
to further speedups alongside enhanced resolution on the noise process.

A separate line of inquiry came from the stochastic simulation community,
where replication -- i.e., multiple simulations with the same input -- is a
common design principle. \citet{ankennman:nelson:staum:2010}'s stochastic
kriging (SK) coupled moment-based empirical variance estimates from
replicates, capturing partially sufficient information about latent variances,
with likelihood-based inference for means.  A downside is that SK requires
{a} minimum degree of replication for all inputs (no fewer
than ten), limiting application. \citet{binois2018practical} merged SK with
ML point inference via maximization.  They showed that fully sufficient
statistics can be used in a completely likelihood-based toolkit, paving a
direct avenue back to fully Bayesian integration.

\subsection{Woodbury likelihood} \label{sec:woodbury}

\citeauthor{binois2018practical}~provided a set of identities relating an
ordinary GP likelihood \eqref{eq:logl}, possibly with additional
heteroskedastic priors \eqref{eq:lamp}, and predictive equations
\eqref{eq:gppred} to ones that involve, potentially, many fewer sufficient
statistics under replication. Explaining these, and ultimately incorporating
them into our Bayesian setting [Section \ref{sec:woodess}], requires an
upgrade in notation.

Let $X_n$ be an $n \times d$ matrix of unique inputs from $X_N$, and let
$a_1, a_2, \dots, a_n$ denote their replication multiplicity in $X_N$.  Let
$\bar{Y}_n = (\bar{y}_1, \bar{y}_2, \dots, \bar{y}_n)$, so that $\bar{y}_i$
stores averages of the $y_i^{(1)}, \dots, y_i^{(a_i)}$-values at each
unique-$n$ input, i.e., $\bar{y}_i = \frac{1}{a_i} {\sum_{j=1}^{a_i}
y_i^{(j)}}$, for $i=1, \dots, n$.  Now, let $\Lambda_n$ denote a diagonal
matrix of latent $\lambda_1, \dots, \lambda_n$ like $\Lambda_N$, but for the
unique-$n$ inputs.  If {homoskedistic} modeling, take $\Lambda_n = g
\mathbb{I}_n$.  \citeauthor{binois2018practical} showed that the following
unique-$n$ log likelihood is identical to the full-$N$ analog
\eqref{eq:logl}, as long as $\Lambda_n$ is repeated by multitude 
$A_n = \mathrm{Diag}(a_1, a_2, \dots, a_n)$ in $\Lambda_N$:
\begin{align}
\log \mathcal{L}(Y_N \mid X_n, \Lambda_n) & \label{eq:wLL}
\propto - \frac{N + \alpha_Y}{2} \log \hat{\tau}^{2}_N - 
\frac{1}{2} \log | \Upsilon_n | 
- \frac{1}{2} \sum_{i=1}^n\left [ (a_i -1) \log \lambda_i + \log a_i \right] \\
\hat{\tau}^{2}_N & = 
\frac{1}{N + \alpha_Y}(Y_N^\top \Lambda_N^{-1} Y_N - \bar{Y}_n^\top A_n \Lambda_n^{-1} \bar{Y}_n + \bar{Y}_n^\top {\Upsilon}_n^{-1} \bar{Y}_n + \beta_Y), \nonumber
\end{align}
where $\Upsilon_n = \Kn +  A_n ^{-1} \Lambda_n$.  The proof involves applying
the Woodbury identity for inverses and determinants
\citep[e.g.,][]{golub:1996}.  Woodbury has been used with GPs before
\citep[e.g.,][]{banerjee2008gaussian, ng2012bayesian}, however this particular
application was new to surrogate modeling and opened up new possibilities for
modeling stochastic simulations.

Compared to the full-$N$ analog this brings two efficiencies: 1) only
cubic-in-$n$ calculations are needed, noting that diagonal $\Lambda_N^{-1}$
and $\Lambda_n^{-1}$ are basic reciprocals and can be stored as vectors; 2) we
only need to learn $n$ latent quantities, not $N$. Notice that $\hat{\tau}^2_N
\ne
\hat{\tau}^2_n \equiv n^{-1}\bar{Y}_n^\top {\Upsilon}_n^{-1} \bar{Y}_n$, which
is proportional to penultimate term in Eq.~\eqref{eq:wLL}. In other words, the
full-$N$ scale is not the same as the unique-$n$ scale. The correction factor,
missing from SK, may be re-expressed as
\begin{equation}
N^{-1} (Y_N^\top \Lambda_N Y_N - \bar{Y}_N^\top A_n \Lambda_n^{-1} \bar{Y}_N) = 
N^{-1} \sum\limits_{i=1}^n \frac{a_i}{\lambda_i} s_i^2
\quad \mbox{ where } \quad 
s_i^2 = \frac{1}{a_i} \sum_{j=1}^{a_i} (y_i^{(j)} - \bar{y}_i)^2 \label{eq:si}
\end{equation}
and can be pre-calculated. Together with $\{\bar{y}_i\}_{i=1}^n$ and $Y_N^\top
\Lambda_N Y_N$, these comprise a complete set of sufficient statistics for
$\Lambda_n$ and any other kernel hyperparameters.

Conditional on estimated quantities, say via MLE as described by
\citeauthor{binois2018practical}, the kriging equations \eqref{eq:gppred} also
benefit from re-expression via Woodbury.
\begin{align}
\mu_n(\mathcal{X}) & = K_{\theta_Y}(\mathcal{X}, X_n)^\top (\Kn + A_n^{-1} \Lambda_n)^{-1} \bar{Y}_n \label{eq:predy} \\
\Sigma_n(\mathcal{X}) & = \hat{\tau}^2_N \left( K_{\theta_Y}(\mathcal{X}, \mathcal{X}) +
\lambda(\mathcal{X})
- K_{\theta_Y}(\mathcal{X}, X_n) (\Kn + A_n^{-1} \Lambda_n)^{-1} {K_{\theta_Y}(X_n, \mathcal{X})} \right) \nonumber
\end{align}
Actually, the top-middle panel of Figure \ref{fig:mcycle}, narrated earlier in
 Section \ref{sec:hetgp}, used these calculations, via replicates.  The bottom
 panel shows $\log (\hat{\tau}_N^2 s_i^2)$-values from the $a_i$ replicates as
 labeled.

\subsection{Woodbury in ESS}
\label{sec:woodess}

\begin{figure}[ht!]
\centering
\includegraphics[scale=0.65, trim=0 0 0 10,clip=TRUE]{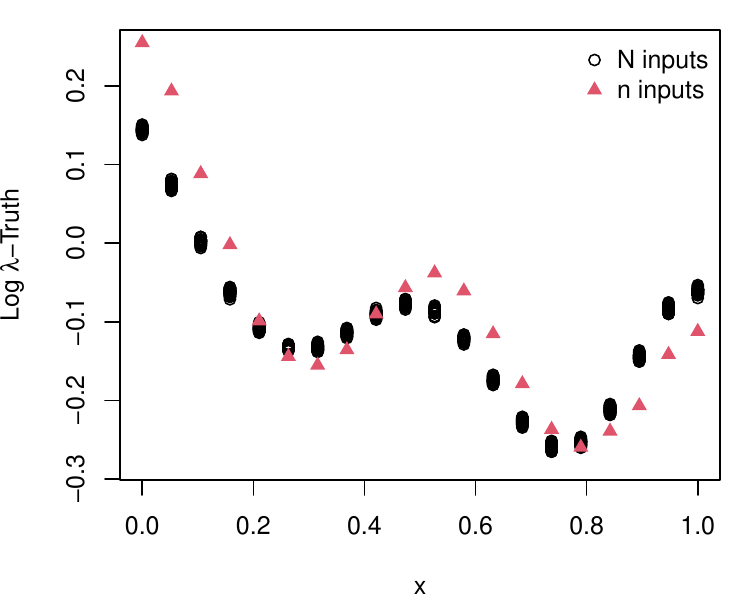}
\includegraphics[scale=0.555, trim=3 15 30 58,clip=TRUE]{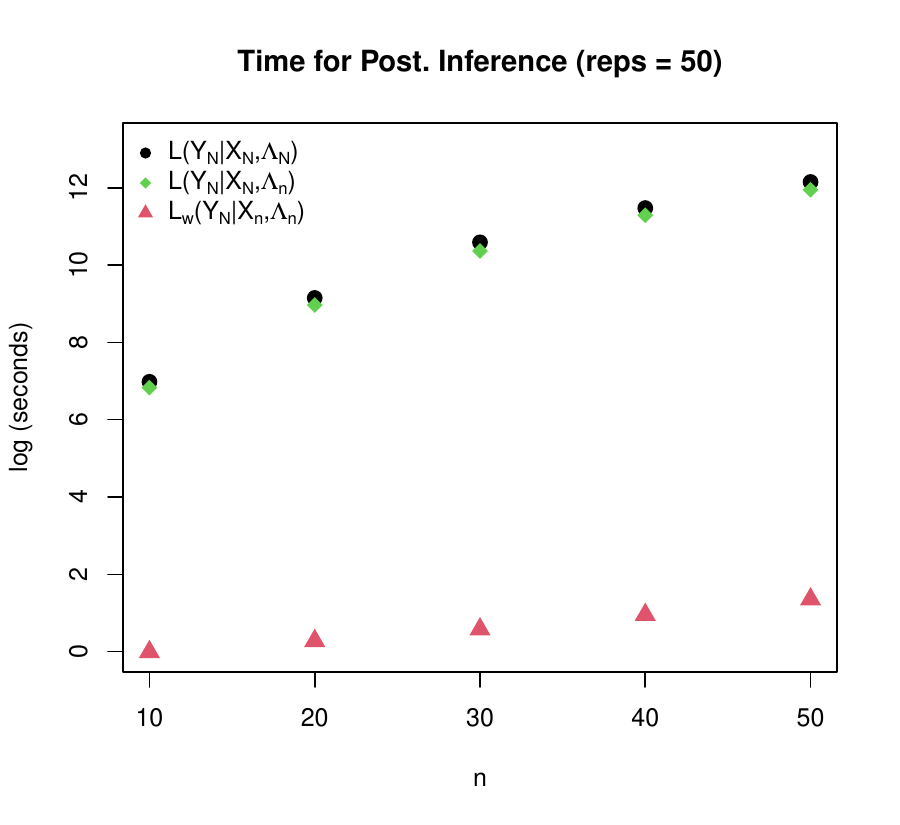}
\caption{{\em Left}: 
exploring statistical with $\Lambda_N$ sampling ({\em black points}) versus $\Lambda_n$
({\em red triangles}). {\em Right}: exploring computational efficiency for increasing
experiment sizes, $N = 50n$ over 1000 MCMC iterations.} 
\label{fig:ess_s3}
\end{figure}

Whereas \citeauthor{binois2018practical}~maximized Eq.~\eqref{eq:wLL} to infer
$\hat{\Lambda}_n$, we propose posterior sampling by ESS
(\ref{eq:ess2}--\ref{eq:ess4}).  Compared to the presentation in Section
\ref{sec:hetgpN}, which used $\Lambda_N$ and the full-$N$ likelihood
\eqref{eq:logl}, a unique-$n$ approach brings two efficiencies: statistical
and computational. These are highlighted in the {\em left} and {\em right}
columns of Figure~\ref{fig:ess_s3}, respectively.  They summarize an
experiment with a simple 1d heteroskedastic process under degree fifty uniform
replication, i.e., $A_n = 50 \mathbb{I}_n$, so that $N = 50n$.

{The {\em left} panel shows the outcome of one (burned-in) ESS
iteration for $\Lambda_N$ and $\Lambda_n$ using $n = 20$ following the setup in
Section \ref{sec:hetgpN}.~} Two totally separate chains were used, and a
residual to the true noise {is} plotted. Since replicates are
un-tethered from one another in the full-$N$ setup (black), they appear
``noisy'' when plotted at the same $X_n$-coordinates.  In the unique-$n$
formulation, there is only one $\lambda_i$-value for each unique input.  This
means lower MC error, i.e., without that ``noise'', both across MCMC
iterations $t$, but also within ESS since the log likelihood(s) \eqref{eq:wLL}
sum over fewer random quantities
\eqref{eq:logl}.

One way around $\Lambda_N$'s statistical inefficiency is to generate
$\Lambda_n$ and replicate it's values $A_n$ times to deduce $\Lambda_N =
\text{Block-Diag}(\lambda_1 \mathbb{I}_{a_1}, \dots, \lambda_n
\mathbb{I}_{a_n})$.  This reduces MC error, but doesn't save much on
computational costs.  See {the} {\em right} panel of Figure
\ref{fig:ess_s3}, offering a timing comparison as $n$ is varied, keeping $N =
50n$.  Black dots involve $\Lambda_N$ sampled directly, whereas green ones use
$\Lambda_N$ deduced from $\Lambda_n$.  Neither is
{competitive with} versions that work only with $X_n$ and
$\Lambda_n$, shown in red.

\subsection{Hierarchical modeling}\label{sec:hier}

Here we complete our Bayesian {\tt hetGP} with a full hierarchical model and
MCMC-based inference for all unknowns. Many of the requisite expressions were
provided earlier, and we shall simply refer to those rather than duplicate
them in order to keep the focus on novel aspects.

\paragraph{(i) Generative model/likelihood:} The most important {\tt hetGP}
components, lying at the top of the hierarchy, are the {data generating/mean
\eqref{eq:hetgp}} and the variance process \eqref{eq:lamp}.  These are
unchanged, except that we use $\Lambda_N = \mbox{Block-Diag}(\lambda_1
\mathbb{I}_{a_1}, \dots, \lambda_n \mathbb{I}_{a_n})$.  When putting these
together, as a marginal log likelihood, we utilize the Woodbury formulation
\eqref{eq:wLL}.  Later, when we discuss {posterior} sampling in paragraph (iii),
we shall need a log ``likelihood'' for the $\Lambda_n$ values only, in 
order to learn the hyperparameters of the variance process.  
Let $\ell_n = \log \mathrm{vec}\, \Lambda_n$ in the following.
\begin{equation}
\log \mathcal{L}(\ell_n  \mid X_n) \propto - \frac{n + \alpha_\lambda}{2} \log \hat{\tau}^{2}_\lambda  - \frac{1}{2} \log | \Knl | \quad 
\mbox{where } \quad \hat{\tau}^2_{\lambda} = \frac{\ell_n^\top {\Knl}^{-1} \ell_n + \beta_\lambda}{n + \alpha_\lambda}  \label{eq:blamLL}
\end{equation} 
under $\tau_\lambda^2 \sim \mathrm{IG}(\alpha_\lambda/2, \beta_\lambda/2)$.
More details for priors are coming next. 

\paragraph{(ii) Hyperparameter priors:} 
Priors for $\tau^2_Y$ and $\tau^2_\lambda$ were covered earlier.
Inverse-Gammas (IGs) are common for scales because they are conditionally
conjugate with Gaussian likelihoods, lending an additional degree of analytic
tractability.  Reference priors ($\tau^2 \sim
\mathrm{IG}(0,0) \propto 1/\tau^2$) yield proper posteriors as long as $n \geq
1$ \citep{berger2001objective}, although we find it helpful to use proper
priors informed by scales that are reasonable after light pre-processing.
Particular prior parameter settings are provided along with other
implementation details in Section \ref{sec:bench}.  Lengthscales $\theta_Y$
and $\theta_\lambda$, each $d$-dimensional, are commonly placed under
independent Gamma priors, e.g., $\theta_Y^{(j)} \stackrel{\mathrm{iid}}{\sim}
G(a, b)$.  Particular $(a,b)$ settings come later, however we note
\citet{binois2018practical} incorporated the restriction that $\theta_Y^{(j)}
< \theta_\lambda^{(j)}$, i.e., that the noise process changes more slowly than
the mean.  We privileged that as an option in our setup. The nugget
of the noise process, $g_\lambda = \epsilon > 0$ can be set to a small,
non-zero value $\epsilon$ for numerical stability.  Estimating a value for
$g$, say under a Gamma prior like the lengthscales, can be helpful if
$\Lambda_n$ is poorly initialized. More in Section \ref{sec:bench}.

\paragraph{(iii) Posterior Sampling:} 

Algorithm \ref{alg:gibbs} summarizes our Metropolis- and ESS-within Gibbs
procedure for sampling from the posterior.  Many of the steps have been
addressed previously, and references to the relevant equations are provided
as comments to the right of the pseudocode.  One exception is for lengthscales.
These require independent, random-walk Metropolis proposals.  Generically
for any $\theta$, of which there are $2d$, we make a positive sliding window
proposal uniformly from half to double the previous value:
\begin{equation}
\theta^\star \sim \mathrm{Unif}\left( \theta^{(t-1)}/2 , 2 \theta^{(t-1)} \right) 
\quad \text{ so } \quad  \label{eq:mh}
\frac{q(\theta^{(t-1)} \mid \theta^\star)}{q(\theta^\star \mid \theta^{(t-1)})} 
= \frac{\theta^{(t-1)}}{\theta^*}
\end{equation}
in the MH acceptance ratio.  The marginal likelihood and prior ratio that's used
depends on which process (means for $Y$ or latent variances $\Lambda_n$), and
is indicated in the comment.  As a shorthand for a vector of partially completed 
$\theta$-values as $t-1 \rightarrow t$, let 
$\theta^{(t \rightarrow)}_j \equiv (\theta_1^{(t)}, \dots, \theta_{j-1}^{(t)},  \theta_j^{(t-1)},
\theta_{j+1}^{(t-1)}, \dots, \theta_d^{(t-1)})$ 
for $j \in \{2,, \dots, d-1\}$ and $\theta^{(t \rightarrow)}_1 =
\theta^{(t-1)}$ and $\theta^{(t \rightarrow)}_d = (\theta^{(t)}_{1:(d-1)},
\theta^{(t-1)}_d)$.

\bigskip

\begin{algorithm}[H]        
\DontPrintSemicolon
Initialize $\theta_{{Y}}^{(1)}, \tl^{(1)}$ and $\Lambda_n^{(1)}$.  \tcp*{See Section \ref{sec:bench}}
\For{$t = 2, \dots, T$}{
    \For{$j = 1, \dots, d$}{
    ${\theta_{j\lambda}^{(t)}} \sim \pi(\theta_{j\lambda} \mid Y_N, {\Lambda_n}^{(t-1)}, X_n, {\theta_{j \lambda}}^{(t\rightarrow)})$ 
    \tcp*{MH (\ref{eq:mh}) via  Eq.\,(\ref{eq:blamLL})}
    $\theta_{jY}^{(t)} \sim \pi(\theta_{jY} \mid {Y}_N, X_n, {\Lambda_n}^{(t-1)}, \theta_{jY}^{(t\rightarrow)})$ 
    \tcp*{MH (\ref{eq:mh}) via  Eq.\,(\ref{eq:wLL})}
    }
    $\log \Lambda_n^{(t)} \sim \pi({\log \Lambda_n} \mid {Y}_N, X_n, \Lambda_n^{(t-1)}, \ty^{(t)}, \tl^{(t)})$
        \tcp*{ESS (\ref{eq:ess3}) via Eqs.\,(\ref{eq:wLL}) \& (\ref{eq:blamLL})}
        $(\hat{\tau}_N^{2(t)}, \hat{\tau}_\lambda^{2(t)}) = \mbox{Latest above}$ 
    \tcp*{via Eqs.\,(\ref{eq:wLL}) \& (\ref{eq:blamLL})}
}
\caption{Metropolis- and ESS-within-Gibbs sampling for Bayesian {\tt hetGP}}
\label{alg:gibbs}
\end{algorithm}

\bigskip

The final step in the algorithm requires some explanation.  Recall that scales
may be analytically integrated out under IG priors, leading to marginal
likelihoods (\ref{eq:wLL} \& \ref{eq:blamLL}). These would have been
calculated in earlier steps as part of ESS or MH acceptance ratios for
proposed/intermediate values.  The quantities involved -- $\hat{\tau}_N^2$ or
$\hat{\tau}_\lambda^2$ -- are not parameters, but sufficient statistics
given samples of other parameters.  You {\em can} obtain samples for
$\tau_N^2$ or $\tau_\lambda^2$ from their IG posterior conditionals, but we do
not require them for anything downstream.  However we {\em do} need
$\hat{\tau}_N^{2(t)}$ or $\hat{\tau}_\lambda^{2(t)}$ for prediction, which can
be recorded before incrementing $t$.

\paragraph{(iv) Prediction:} 
After eliminating any burn-in, etc., let $\{
\tl^{(t)}, \ty^{(t)}, \Lambda_n^{(t)}, \tau_\lambda^{2(t)}, \tau_N^{2(t)} \mid
t \in \mathcal{T} \}$ denote the samples saved from Algorithm \ref{alg:gibbs}.
Predictions $Y(\mathcal{X})^{(t)}$ at new inputs $\mathcal{X}$ requires
$\Lambda(\mathcal{X})^{(t)}$. Since these depend estimated scales $\{
\hat{\tau}_\lambda^{2(t)}, \hat{\tau}_{{N}}^{2(t)} \}$, both
$\ell(\mathcal{X}) \equiv \log \mathrm{vec}\, \Lambda(\mathcal{X})$ and
$Y(\mathcal{X})$ are conditionally Student-$t$ with $n-1$ degrees of freedom.
{When $n \gg 100$, as is common, we} feel comfortable with a
Gaussian approximation as follows. Use Eq.~\eqref{eq:gppred} with
$Y_N \equiv \log \mathrm{vec}\, \Lambda_n^{(t)}$ and $K \equiv
K^{(t)}_{\theta_\lambda}(X_n)$ to obtain $\mu^{\ell(t)}_n$ and
$\Sigma^{\ell(t)}_n$, then $\ell (\mathcal{X})^{(t)} \sim \mathcal{N}
(\mu^{\ell(t)}_n, \Sigma^{\ell(t)}_n)$. Take $\Lambda(\mathcal{X})^{(t)}
= \mathrm{diag} \exp\{\ell(\mathcal{X})^{(t)}\}$ along with $\{ \ty^{(t)}, \tau_N^{2(t)} \}$
in Eq.~\eqref{eq:predy} yielding $\mu^{\mathcal{Y}}_n(\mathcal{X})^{(t)} $ and
$\Sigma^{\mathcal{Y}}_n(\mathcal{X})^{(t)}$. Although these could provide samples of
$Y(\mathcal{X})^{(t)}$, it is usually more expedient to summarize moments.
Dropping $\mathcal{X}$ for compactness, we use
\begin{equation}
\bar{\mu}^\mathcal{Y}_n =  \frac{1}{|\mathcal{T}|} \sum_{t \in \mathcal{T}} \mu^{\mathcal{Y}(t)}_n  \quad \text{and } \quad  \label{eq:lawExp}
\bar{\Sigma}^\mathcal{Y}_n =  \frac{1}{|\mathcal{T}|} \sum_{t \in \mathcal{T}} \Sigma^{\mathcal{Y}(t)}_n + 
\frac{1}{|\mathcal{T}| - 1} \sum_{t \in \mathcal{T}} ( \mu^{\mathcal{Y}(t)}_n -  \bar{\mu}^\mathcal{Y}_n) ( \mu^{\mathcal{Y}(t)}_n-  \bar{\mu}^\mathcal{Y}_n)^\top,
\end{equation}
where the latter calculation for $\bar{\Sigma}^\mathcal{Y}_n$ involves an application of
the law of total variance.  When $\mathcal{X}$ is large, i.e., large $n'$,
storage and decomposition for a large $n' \times n'$ matrix
$\Sigma(\mathcal{X})^{(t)}$ can be cumbersome and represent overkill.  For
many applications it is sufficient to save only variances along the diagonal,
which may be calculated point-wise and potentially in parallel over all $x \in
\mathcal{X}$.  For example, the error-bars plotted in the {\em top-right} panel
of Figure \ref{fig:mcycle} require only point-wise variances.

\section{Large simulation campaigns}\label{sec:vec}

When $N \gg n$ the Woodbury likelihood can make an intractable inferential
setting computationally manageable. For example, the slowest case in the {\em
right} panel of Figure \ref{fig:times} involves $(n,N)=(50, 2500)$, which
takes two days to run on full-$N$ calculations but fewer than ten seconds with
unique-$n$ ones.   However, a large number of unique inputs, $n$, can still be
a bottleneck. The NOAA-GLM campaign introduced in Section \ref{sec:intro},
with $n > 300{,}000$ and $N=30n$, is a non-starter.

There are recently many sparse and low-rank approximations to GP inference and
prediction
\citep[e.g.,][]{cressie2008fixed,katzfuss2011spatio,emery2009kriging,gramacy2015local,
cole2021locally,furrer2006covariance,kaufman2008covariance,stein2013statistical,datta2016hierarchical,stein2004approximating,titsias2010bayesian}.
More in \citet[][Chapter 9]{gramacy2020surrogates}.  However, none have been
adapted to {\tt hetGP}s. \citet{holthuijzen2024synthesizing} custom-built an
SK-like approach so that software for a variant of the Vecchia approximation
\citep{katzfuss2022scaled}, which we shall review momentarily, could be used.
However, their setup precluded linking mean and variances processes. In spite
of that, the result was an impressive new capability, but also one ripe for
improvement.

\subsection{Vecchia approximation}\label{sec:vecR}

\citet{vecchia1988estimation}'s idea was ahead of its time, and has recently
seen a resurgence of interest \citep[e.g.][]{datta2022nearest,
katzfuss2020vecchia,stroud2017bayesian,sauer2023vecchia} with modern advances
in hardware architecture and sparse matrix libraries. The basic idea stems
from an elementary identity allowing joint probabilities to be factorized as a
product of cascading conditionals.  Here we express that identity for the
likelihood, with our GP application in mind, and privilege the unique-$n$
representation for reasons that will be discussed shortly.
\begin{equation}\label{eq:cond}
\mathcal{L}(Y_n) = \prod\limits_{i=1}^{n} \mathcal{L}(y_i \mid Y_{c(i)}),
 \quad \mbox{ where } \quad c{(i)} = \{1, 2, \dots ,i-1 \}
\end{equation}
is the so-called conditioning set, and $Y_{c(i)} \equiv \{y_i : i \in c(i)\}$.
This identity is true for any ordering of indices.  Dropping some of them,
i.e., $c{(i)} \subset \{1, 2, \dots, i-1\}$, yields an approximation whose
quality depends both on the indexing and their composition in $c(i)$. Although
there are many choices here, we follow the the suggestion of recent studies
(see cites above) and use random indexing and at most $m$
{Euclidean} nearest neighbors (NN) for $c(i)$, i.e., $| c(i) |
= \min(m, i-1)$.

This cascade \eqref{eq:cond} is perfect for GPs because 
each conditional is a GP prediction, e.g.,
Eqs.~\eqref{eq:gppred} or
\eqref{eq:predy}.  Specifically ${\mathcal{L}(y_{i} \mid
Y_{c(i)}) \equiv \mathcal{N}_1 (\mu_i , \sigma_i^2)}$, where
\begin{align}
\mu_i & = B_i Y_{c(i)} & \label{eq:uni}
B_i & = \Sigma(x_i, X_{c(i)}) \Sigma(X_{c(i)})^{-1} \\
\sigma_i^2 &= \Sigma(x_i) - B_i \Sigma(X_{c(i)}, x_i).
\nonumber
\end{align}
Notice that each $\Sigma(X_{c(i)})^{-1}$ is at worse $m \times m$, limiting
computation to flops in $\mathcal{O}(n m^3)$, which represents a potentially
dramatic improvement if $m \ll n$.

\citet{katzfuss2021general} go further to show how the full precision matrix
implied by the approximation may be represented as a sparse Cholesky
factor $\Sigma(X_n)^{-1} \approx \Uxn\Uxtn$, where $\Uxn$ is an upper
triangular matrix whose entries may be populated in parallel using
Eq.~(\ref{eq:uni}):
\begin{equation}\label{eq:Umat}
\Uxn^{ij} = \begin{cases}
         \frac{1}{\sigma_i} \qquad & i = j\\ 
        - \frac{1}{\sigma_i} B_i[\text{index of $j$ in $c(i)$}] \qquad & j \in c(i) \\
        0 \qquad & \text{otherwise.}
    \end{cases}
\end{equation} 
Cholesky solves required for inversion and determinant are fast when $\Uxn$ is
stored and manipulated with modern sparse matrix libraries. More details on
our own implementation, borrowed liberally from \citet{katzfuss2021general},
are provided in Section \ref{sec:bench}.

When predicting, the Vecchia sparse inverse structure may be extended to a
stacked covariance \eqref{eq:stack} with testing inputs $\mathcal{X}$. That
is,  $\Sigma_\text{stack} \approx (U_{\text{stack}}
U_\text{stack}^\top)^{-1}$.  This requires extending conditioning sets from
$n$ to $n+n_p$ indices.  We follow \citet{sauer2023vecchia}, preserving
indices and conditioning sets used for training, and randomly assigning
indices from $\{n+1, \dots, n + n_p\}$ for $\mathcal{Y}(\mathcal{X})$ . Then
$U_{\text{stack}}$ can be built using Eq.~\eqref{eq:Umat} and $c(i)$ via NN as
usual. In this way, you never condition training on testing quantities. With
components of $U_{\text{stack}}$ partitioned as follows
\begin{align} 
    U_{\text{stack}} = \label{eq:Ustack}
    \begin{bmatrix} 
         U_n & U_{n, \mathcal{X}}\\
            0 &  U_\mathcal{X}\\ 
    \end{bmatrix}
        \;\; \mbox{ s.t.~} \;\;
        \Sigma_\text{stack} \approx (U_{\text{stack}}U_\text{stack}^\top)^{-1} 
        =
        \begin{bmatrix} 
            U_n U_n^\top + U_{n, \mathcal{X}} U^\top_{n, \mathcal{X}} & U_{n, \mathcal{X}} U^\top_\mathcal{X}\\
            U_\mathcal{X} U^\top_{n, \mathcal{X}} &  U_\mathcal{X} U^\top_\mathcal{X}\\ 
        \end{bmatrix}^{-1},
\end{align}
the kriging equations may be characterized as $\mathcal{Y}(\mathcal{X}) \mid
Y_N, X_n \sim \mathcal{N} \left(\mu^\mathcal{Y}_n(\mathcal{X}),
\Sigma^\mathcal{Y}_n(\mathcal{X}) \right)$ with
\begin{equation}
\mu^\mathcal{Y}_n (\mathcal{X}) = \-(U^\top_\mathcal{X})^{-1} U^\top_{n, \mathcal{X}} {Y}_n 
\quad \mbox{ and } \quad 
\Sigma^\mathcal{Y}_n(\mathcal{X}) = (U_\mathcal{X} U_\mathcal{X}^\top )^{-1} \label{eq:predvec}.
\end{equation}
This may still computationally intensive when $n_p = |\mathcal{X}|$ is large.
Often the full joint predictive is overkill -- cross terms in
$\Sigma_n^{\mathcal{Y}}$ are not required -- and only the diagonal (variance)
terms are needed. In such settings, independent applications of
Eq.~\eqref{eq:predvec}, treating each $x \in \mathcal{X}$ as a singleton
$\mathcal{X}$, potentially in parallel, can be more expedient.

\subsection{Proof of concept} \label{sec:vecN}

Before delving into the details of how we situate Vecchia within our Bayesian
{\tt hetGP} framework, we provide an empirical analysis that tests potential
in our setting.  Vecchia {\em can} be generic, providing a sparse inverse
Cholesky for any MVN with a distance-based kernel.  Existing software could,
for example, be deployed un-altered in the full-$N$ setup (conditional on
$\Lambda_N$) in both MLE \citep{katzfuss2022scaled} and Bayesian
\citep{sauer2023vecchia} settings. What happens if we use that in a heavy
replication setting? Conversely, what is the value of deploying Vecchia on the
Woodbury likelihood \eqref{eq:wLL} instead? Our presentation coming shortly in
Section \ref{sec:vecn} is intricate.  Is it worth it?

\begin{figure}[ht!]
\centering
\includegraphics[scale = 0.51, trim=5 65 25 15,clip=TRUE ]{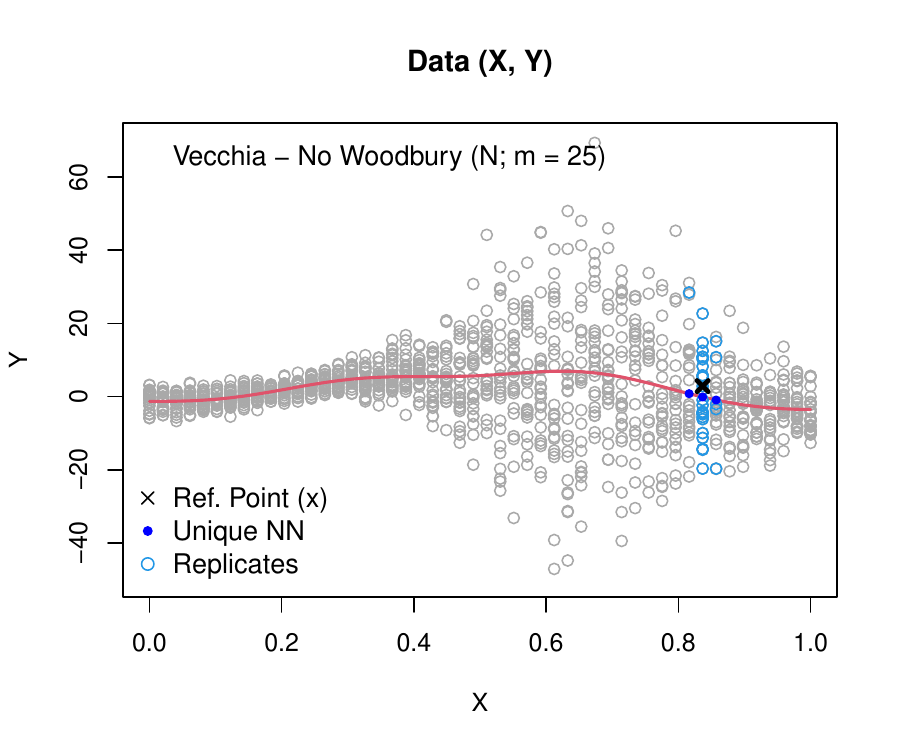}
\includegraphics[scale = 0.51,trim=5 65 25 15,clip=TRUE]{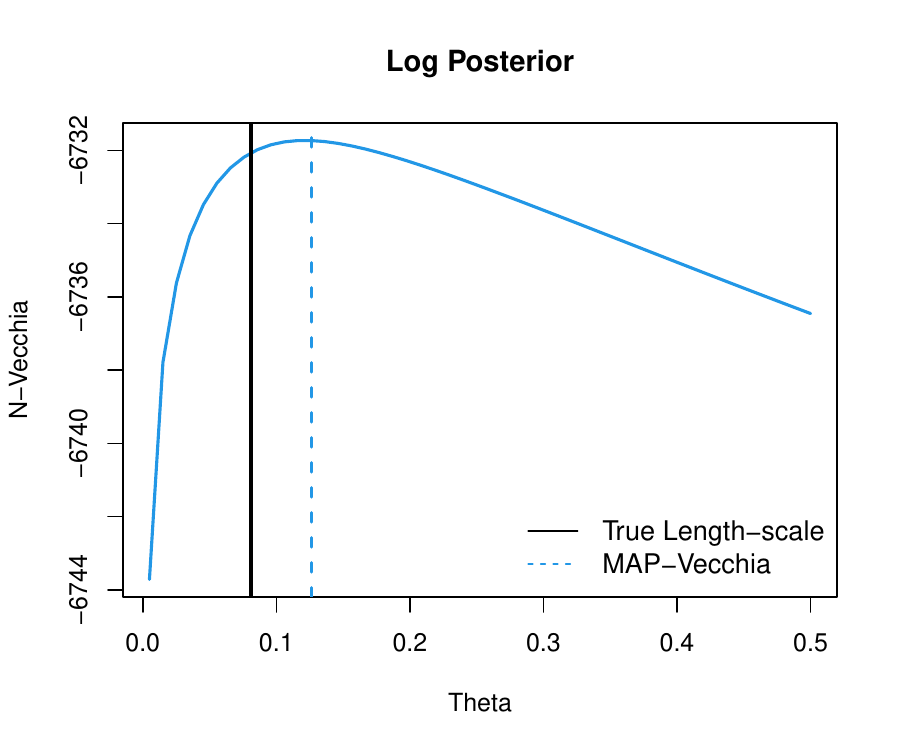}
\includegraphics[scale = 0.51,trim=30 65 20 15,clip=TRUE]{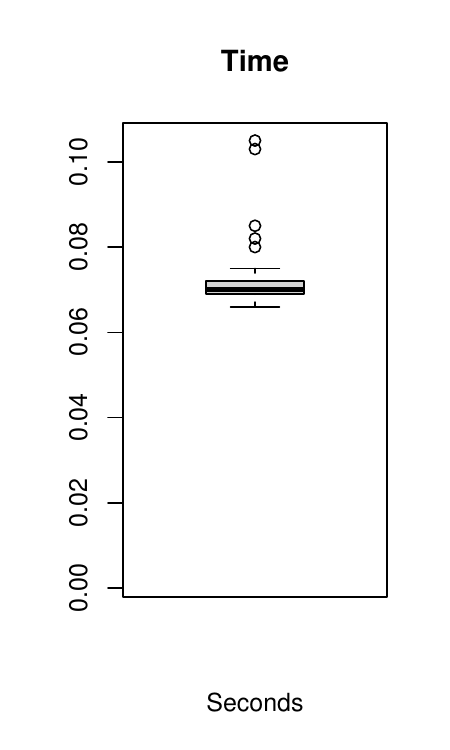}\\
\includegraphics[scale = 0.51,trim=5 0 25 50, clip=TRUE]{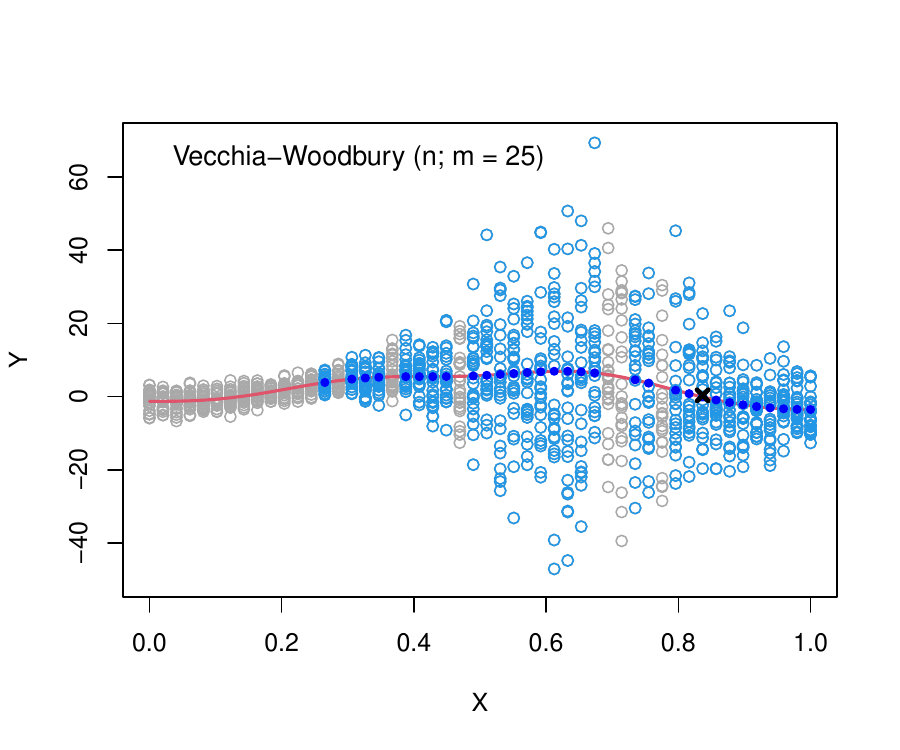}
\includegraphics[scale = 0.51,trim=5 0 25 50, clip=TRUE]{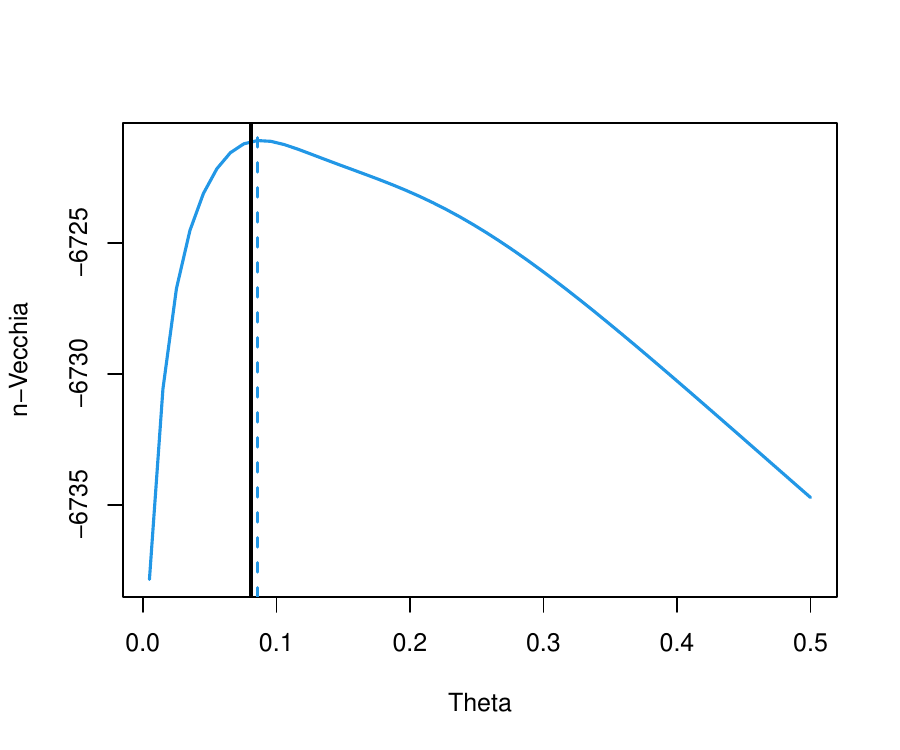}
\includegraphics[scale = 0.51,trim=30 0 20 55, clip=TRUE]{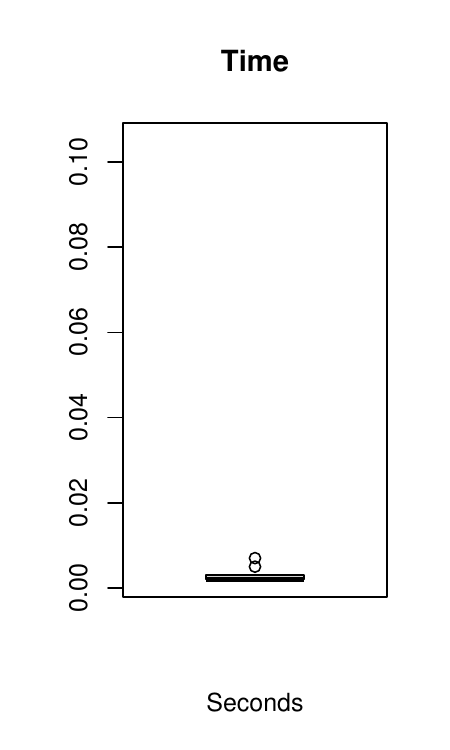}
\caption{{\em Left:} Vecchia approximation showing chosen NN {without ({\em top}) and
with ({\em bottom}) Woodbury likelihood}; {\em middle:} posterior surfaces and MAP
estimate; {\em right} computation time from fifty reps.
\label{fig:vec_s4}}
\end{figure}

Figure \ref{fig:vec_s4} considers the simple case of evaluating the likelihood
for a lengthscale $\theta$ in a simple 1d test case where $N = 25n$. The top
row corresponds to Vecchia with an ordinary, full-$N$ likelihood, whereas the
bottom is with the unique-$n$ Woodbury likelihood.  Both use $m=25$, random
indexing and NN. Consider each column in turn.  The {\em left} column shows
the neighborhood involved in a particular row of the inverse Cholesky factor,
$U_N$ ({\em top}) or $U_n$ ({\em bottom}).  With $U_N$ there is low diversity.
The selected $m = 25$ points span less than 5\% of all training pairs. Under
this particular random indexing there are only three unique inputs in the
conditioning set. Without uniform replication, fixed indexing and {\em a
priori} known degree, you can't predict how many unique inputs will be in a
$U_N$ conditioning set.  With $U_n$, where replicates are nearly ``free''
under Woodbury, we get exactly $m$ by design, so the neighborhood spans nearly
half the data.

The {\em middle} panels show the approximated posterior surface for $\theta$.
Notice how the unique-$n$ one is more peaked around the true value.  Finally,
consider compute time in the {\em right} column.  The only difference here is
$N$ v.~$n$, since both involve the same $m$ and thus the same
$\mathcal{O}(m^3)$ decomposition cost.  It's just a question of how many
decompositions are required for sampling via MCMC or  {to} find an MLE.  The unique-$n$
version does fewer of these, by $n/N$, so it is consistently faster.

\subsection{Vecchia--Woodbury modeling and inference}
\label{sec:vecn}

Our full {Vecchia}-approximated {\tt hetGP} may be expressed as follows.
\begin{align}
Y_N & \sim \mathcal{N} \left( 0 , \tau^{2}_N  \UNinv \right) && \mbox{ where } &
\UNinv &\approx \KN + \Lambda_N \label{eq:bhetgpVec}\\
\log \mathrm{vec}\, \Lambda_n & \sim \mathcal{N} \left( 0 , \tau^{2}_\lambda  \Uxlinv \right)
&& \mbox{ and } & \Uxlinv &\approx \Knl + g \mathbb{I} \nonumber
\end{align}
Figure \ref{fig:vec_s4} demonstrated that full-$N$ Vecchia does not handle
replicates well. So instead we prefer {to} build a sparse Cholesky factor for
$\Upsilon_n \approx
\Uxninv$ using Eq.~\eqref{eq:Umat} with the following quantities: 
\begin{align}
B_i & = K_{\ty} (x_i, X_{c(i)}) \Upsilon_{c(i)}^{-1} && \mbox{ where } & \label{eq:Uhet}
\Upsilon_{c(i)} = K_{\tl} (X_{c(i)}) + A_{c(i)}^{-1} \Lambda_{c(i)} \\
\sigma_i^2 & = K_{\ty}(x_i) + \lambda(x_i) - B_i K_{\ty}(X_{c(i)}, x_i).
\nonumber
\end{align}
$\Uxl$ involves a straightforward application of Eq.~(\ref{eq:Umat}) with
$\sigma_i$, $B_i$ defined in Eq.~(\ref{eq:uni}) using $\Sigma(\cdot) =
K_{\tl} (\cdot)+ g \mathbb{I}$.   We share a NN conditioning set and random
ordering across $\Uxninv$ and $\Uxlinv$, i.e., identical $X_{c(i)}$
but different hyperparameters.  Other details follow.

\paragraph{(i) Prior:}
Sampling $\log \mathrm{vec}\, \Lambda_n^\star$ for ESS \eqref{eq:ess2}
requires an $n$-dimensional MVN draw for $\log \mathrm{vec}\, \Lambda_n^{\mathrm{prior}}$.
While an expensive cubic operation for dense $\Sigma$, this is fast via
Vecchia and $\Uxn$ or $\Uxl$.  For example, first sample $Z_n \sim
\mathcal{N}(0, \mathbb{I}_n)$ and then solve the sparse system $U_x^\top \ell =
Z_n$ for $\ell$. Save $\log \mathrm{vec}\, \Lambda_n^{\mathrm{prior}}
\leftarrow \ell$.

\paragraph{(ii) Likelihood:}
ESS and Metropolis steps [Algorithm \ref{alg:gibbs}] require $\mathcal{L}(Y_N
\mid X_n, \Lambda_n)$ and $\mathcal{L}(\log \Lambda_n \mid X_n)$. First, adjust
Eq.~\eqref{eq:wLL} to use sparse $\Uxn$ in place of $\Upsilon_n$.
\begin{align}
  	\log \mathcal{L}_v(Y_N \mid X_n, \Lambda_n) & \propto  \nonumber
   \log  |\Uxninv|^\frac{1}{2} - \frac{N + \alpha_Y}{2} \log \hat{\tau}_N^2 
   - \frac{1}{2} \sum\limits_{i=1}^{n}[(a_i - 1)\log \lambda_i + \log a_i]\\
 & \propto \sum\limits_{i=1}^{n} \log \Uxn^{ii}
 - \frac{N + \alpha_Y}{2} \log \hat{\tau}_N^2 - 
 \frac{1}{2} \sum_{i=1}^{n}[(a_i - 1)\log \lambda_i + \log a_i], \label{eq:loglvec} \\ 
\mbox{where } \quad \hat{\tau}_N^2 & \approx \frac{1}{N + \alpha_Y} \left[ \sum\limits_{i=1}^n \frac{a_i}{\lambda_i} s_i^2 + \bar{Y_n}^\top (\Uxn \Uxtn) \bar{Y}_n + \beta_Y \right]  \nonumber
\end{align}
Replacing $\Knl$ calculations \eqref{eq:blamLL} with sparse $\Uxl$ ones 
is straightforward.  Using $\ell_n \equiv \log \mathrm{vec}\, \Lambda_n$, 
$$
\log \mathcal{L}_v(\Lambda_n \mid X_n) 
\propto \sum_{i=1}^{n} \log {\Uxl}^{ii}  - \frac{n + \alpha_\lambda}{2} \log \taul \label{eq:loglv}
\quad \mbox{ where } \quad
 \taul = \frac{ \ell_n^\top \Uxlinv \ell_n + \beta_\lambda}{n + \alpha_\lambda}. \nonumber
$$

\paragraph{(iii) Prediction:}

$\mathcal{Y}(\mathcal{X})$ requires $\Lambda(\mathcal{X})$,
available via $\ell(\mathcal{X}) \equiv \log \mathrm{vec}\
\Lambda(\mathcal{X})$ and Eq.~\eqref{eq:predvec}. I.e., an ordinary Vecchia
prediction for the latent variance. First construct
$U^{(\lambda)}_{\text{stack}}$ via Eqs.~(\ref{eq:uni}--\ref{eq:Umat}),
 then partition \eqref{eq:Ustack} into $\Uxl$ and $\Uell$.
Finally, $\ell(\mathcal{X}) \mid Y_N, X_n \sim \mathcal{N} \left(\mu_n^\ell(\mathcal{X}), \Sigma_n^\ell(\mathcal{X}) \right)$, where
\begin{equation}
\mu_n^{\ell}(\mathcal{X}) = \-(\Uellt)^{-1} \Uxell \ell_n\
\quad \mbox{ and } \quad \label{eq:lamvecpred}
\Sigma_n^{\ell}(\mathcal{X})  = \taul \left( \Uell \Uellt \right)^{-1}.
\end{equation}
When $n_p$ is large, it may be overkill to sample the latent quantity
$\ell(\mathcal{X})$. \citet{sauer2023vecchia} argue that it is often
sufficient to simply use the mean $\mu_n^\ell$ in lieu of $\ell$.  Not needing
$\Sigma_n^{\ell}$ allows a pointwise approach, as discussed around
Eq.~\eqref{eq:predvec}, which is faster.  However,
\citeauthor{sauer2023vecchia}'s latents were warped inputs and ours are (log)
variances.  Instead of taking $\Lambda(\mathcal{X}) =
\exp\{\ell(\mathcal{X})\}$, we follow \cite{holthuijzen2024synthesizing} and
use $\Lambda(\mathcal{X}) = \exp\{\mu_n^{\ell}(\mathcal{X}) + \Phi^{-1}_{0.95}
\sigma_n^{\ell}(\mathcal{X})\}$ which can be calculated pointwise since it only
requires $\sigma_n^{2\ell} = \mathrm{diag}(\Sigma_n^{\ell})$. Given
$\Lambda(\mathcal{X})$, prediction follows
Eqs.~(\ref{eq:Ustack}--\ref{eq:Uhet}) using $(\bar{Y}_n, \tauy)$ from
Eq.~\eqref{eq:loglvec}. We have $ Y(\mathcal{X}) \mid Y_N, X_n \sim
\mathcal{N} \left(\mu_n^\mathcal{Y}(\mathcal{X}),
\Sigma_n^\mathcal{Y}(\mathcal{X}) \right)$ where
$$  
\mu_n^\mathcal{Y}(\mathcal{X}) = \-(U^\top_\mathcal{X})^{-1} U^\top_{n, \mathcal{X}} \bar{Y}_{n} 
\quad \mbox{ and } \quad 
\Sigma_n^\mathcal{Y}(\mathcal{X}) = \tauy (U_\mathcal{X} U_\mathcal{X}^\top )^{-1}.
$$


\subsubsection*{An illustration}

\begin{figure}[ht!]
    \centering
    \includegraphics[scale = 0.65, trim=2 0 5 0, clip = TRUE]{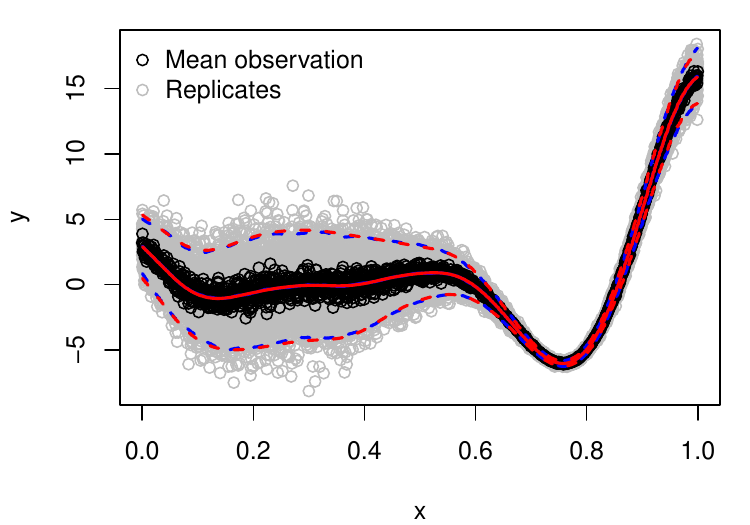}
    \includegraphics[scale = 0.65, trim=2 0 5 0, clip = TRUE]{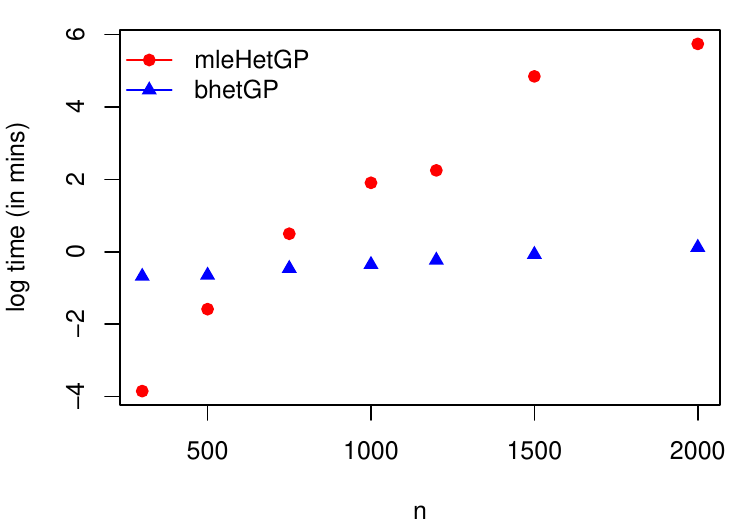}
    \caption{{\em Left:} predictions on 1D toy example using ordinary {\tt
    hetGP} and our {Bayesian/Vecchia hetGP} for $n = 2000$ and
    {$N=10n$}. {\em Right:} time in log minutes varying $n$}
    \label{fig:times}
\end{figure}

Here we explore how a Bayesian {\tt hetGP} with Vecchia compares to an
ordinary (MLE/non-Vecchia) {\tt hetGP} using an illustrative example
introduced by \citet{binois2018replication}.  The setup is $Y(x) = f(x) +
\varepsilon$ for $\varepsilon \sim \mathcal{N}(0, r(x))$ where ${f(x) = (6x -
2)^2 \sin(12x - 4)}$ and  $r(x) = 1.1 + \sin(2 \pi x)$. We consider varying
$n$ using $X_n$ from an LHS in $[0,1]$ and tenfold ($N=10n$) replication.  The
{\em left} panel of Figure \ref{fig:times} shows the data and fits when
$n=2000$. Observe that both {\tt hetGP}s give about the same estimates -- the
lines are on almost top of one another.

The {\em right} panel shows that our Bayesian/Vecchia version is much faster.
{The timings reported include pre-processing steps
(e.g., initialization, finding replication, etc.), fitting, and predictions 
for both methods}. To level the playing field -- MCMC samples versus MLE optimization iterations
-- we limited both to 500 pairs of likelihood evaluations (one for each
process). Sometimes the MLE converged in fewer iterations. The $n=2000$
setting is right on the edge of what's manageable without Vecchia. Whereas
it's clear that the Bayesian/Vecchia alternative would be comfortable with
larger $n$, entertained next. {We note that an MLE-{\tt
hetGP} with Vecchia might be a better match in large $n$ settings, in terms
of comparing apples to apples.  However, no such technology exists.  
Adapting Vecchia to an optimization context, 
requiring additional derivative information, is non-trivial.
By contrast, Vecchia situates nicely within a Bayesian/ESS scheme where it
 enjoys additional UQ benefits.}

\section{Implementation and benchmarking} \label{sec:bench}

Here we provide implementation details and outline our framework for
empirical assessment on benchmark exercises. These include a classic
stochastic queuing problem and our motivating NOAA-GLM simulations. In what
follows, we refer to our method as {\tt bhetGP}. Throughout we use Woodbury
and Vecchia (with $m=25$) without fanfare.  
{We benchmark against} an ordinary {\tt hetGP} via the CRAN
package, when possible, and \citet{holthuijzen2024synthesizing}'s SK-like
alternative, which is tailored to the NOAA-GLM campaign.  {A
referee recommended comparing to stochastic variational sparse inference
heteroskedastic Gaussian processes \citep[SVSHGP;][]{liu2020large}.  However,
we found that their {\sf Python} library was not capable of handling the scale
of experiments presented here. Appendix \ref{app:py} provides a comparison on
the 1D toy data from Section \ref{sec:vecn}.}

\subsection{Implementation}

In many respects, {\tt bhetGP} is a hybrid of the original {\tt hetGP} and
\citet{sauer2023vecchia}'s deep GP. Our {\sf R} package {\tt bhetGP} on CRAN
\citep{bhetGP} is a hybrid of {\tt hetGP} and {\tt deepgp} \citep{deepgp}
packages.\footnote{{You might think that this means that {\tt
bhetGP} is more complex than {\tt hetGP}, but surprisingly not. ESS sampling
is easier than MLE optimization because the latter requires gradients which
are a hefty implementation lift. Our {\tt bhetGP} is involves about 2/3 as
much {\sf R} and {\sf C++} code as {\tt hetGP}.}} {\tt OpenMP}, {\tt Matrix}
\citep{Matrix} and {\tt RcppArmadillo}
\citep{Rcpp,eddelbuettel2014rcpparmadillo} are used to manipulate $\Uxn$ and
$\Uxl$, and predictive analogues.  All examples presented herein are
reproducible using code from our Git repo:
\url{https://bitbucket.org/gramacylab/bhgp/src/main/examples/}.

Our priors are chosen to be vague for inputs $X_n$  and $Y_n$ that are,
ideally, {pre-processed into sensible ranges like $[0, 1]^d$
and $[-2,2]$}, respectively, with adjustments to adapt to irregularities. For
example, all lengthscales $\theta_{[\cdot]}$ use a $G(a=1.5, b)$ prior where
$b$ is set based on the maximum squared distance in $X_n$, following
\cite{gramacy2016lagp}. We adapt \citet{binois2018practical}'s rule and
enforce $\theta_\lambda > \theta_Y$, separately in each coordinate, via the
prior.  This encodes an belief that the variance changes slower than the mean.
We use proper priors for scales $\tau^2, \tau^2_\lambda
\stackrel{\mathrm{iid}}{\sim}\mathrm{IG}( a/2, b/2)$ with $(a,b)=(10, 4)$,
although other settings and improper $(a,b)\equiv (0,0)$ are also supported.
All defaults are adjustable, although we hold them fixed at values reported
here throughout our empirical work. {When using Vecchia,
Euclidean NNs are determined on pre-processed inputs, as described above.}

We use a total of $1{,}000$ MCMC iterations, discarding 500 as burn-in and
thinning by ten so that all predictions are based on $T=50$ samples.  We chose
such limited MCMC for two reasons: (1) to serve as a testament to the
excellent mixing provided by ESS for $\Lambda_n$; and (2) so our
{MC} experiments, with many repetitions, didn't take too long.  Throughout, we
use pointwise prediction which is fast and parallelized via {\tt foreach}
\citep{foreach} in {\sf R}, and aggregate moments via Eq.~\eqref{eq:lawExp}.
We follow \citet{sauer2023vecchia}, and others, to use a larger conditioning
set of size $m' = 200$ for prediction, which is still very fast with just
$T=50$ samples. 

Although MCMC mixing is excellent, convergence can be slow when $n$ is large
and the starting latent variances $\Lambda_n^{(0)}$ are chosen poorly. When $n
< 500$ or so it doesn't matter {what} $\Lambda_n^{(0)}$ is as long as it's
smooth.  (A ``noisy'' $\Lambda_n^{(0)}$ is problematic when $g_\lambda =
\epsilon$, in which case you could choose to estimate $g$.  
{While this helps cope with a ``jagged'' initialization in
early MCMC iterations, it helps to nudge $g^{(t)} \rightarrow 0$ during
burn-in to ensure an estimated noise process that is ultimately smooth.} This
is a case discussed at length by \citet{binois2018practical},
{who provide a lemma explaining that the MLE is at $g = 0$}.
Our software supports both {options}, but we prefer $g_\lambda
= \epsilon$ and a thoughtfully chosen, smooth $\Lambda_n^{(0)}$.)
{Another} option is constant $\Lambda_n^{(0)} = c$ where $c$
is chosen as 10\% of the estimated marginal variance of the $y_i$ values.

We prefer a more adaptive initialization in all situations, but especially
with large $n$. Basically, we perform pre-fits of the data from other
software, which either provides $\Lambda_n$ directly (e.g., {\tt hetGP} when
$n$ is small), or provides residuals whose sums of squares can be smoothed
into predicted $\Lambda(\mathcal{X})$.  Such fits can also furnish initial
lengthscale and scale estimates.  We focus our brief description here on
{the} more interesting, large-$n$ case where our approach to
initialization is inspired by \citet{holthuijzen2024synthesizing}'s use of
scaled Vecchia
software.\footnote{\url{https://github.com/katzfuss-group/scaledVecchia}}
First fit unique-$n$ data $(X_n,
\bar{Y}_n)$. Then calculate $s_i$ values as in Eq.~(\ref{eq:si}) but with a
predicted $\hat{y}_i$ instead of $\bar{y}_i$. This is important when $a_i$ are
small. If all $a_i$ are large $\bar{y}_i$ is fine; a pre-fit isn't strictly
necessary for $s_i$. Finally, a second scaled Vecchia is fit to $\log s_i$
values, being careful to adjust for scale (i.e., subtract off $\log
\hat{\tau}^2_s$) before assigning $\log \mathrm{vec}\, \Lambda_n^{(0)}$.
{This strategy results in a quick, informed estimate for the
latent noise process to start the MCMC.  It works even for large examples such
as our motivating lakes problem ($\approx$ 9 million observations).}

We are ready now to turn to our benchmarking exercises.  Our primary metrics,
both measured out of sample, are root mean square error RMSE, where lower is
better, and proper score \citep[Eq.~(27) in][]{gneiting2007strictly}, where
higher is better.  



\subsection{Assemble-to-order (ATO)}

ATO is an inventory problem wherein the goal is to optimize a daily profit
output based on number of products sold \citep{hong2006discrete}.  The
simulator \citep{xie2012assemble} is implemented in {\sf MATLAB}, which we
access through {\tt R.matlab} \citep{R.matlab}, and involves random demand for
products whose parts have random procurement times.  The version we study here
involves five products built from eight items whose stock ($d$ columns of
inputs $X_n$) ranges in $\{1, 2 \dots 20\}$.  Our experimental setup mirrors
\cite{binois2018practical}, with a slight variation to entertain larger $n$.
Specifically, we conducted a MC experiment with random training and testing
sets composed of mutually exclusive subsets of a master simulation campaign
performed offline.  That campaign involved uniform $X_{n_{\mathrm{tot}}}$ with
$n_{\mathrm{tot}}=12{,}000$ and tenfold replication, so that $N=120{,}000$.

\begin{figure}[ht!]
    \centering
    \includegraphics[scale = 0.5, trim=0 5 5 2, clip = TRUE]{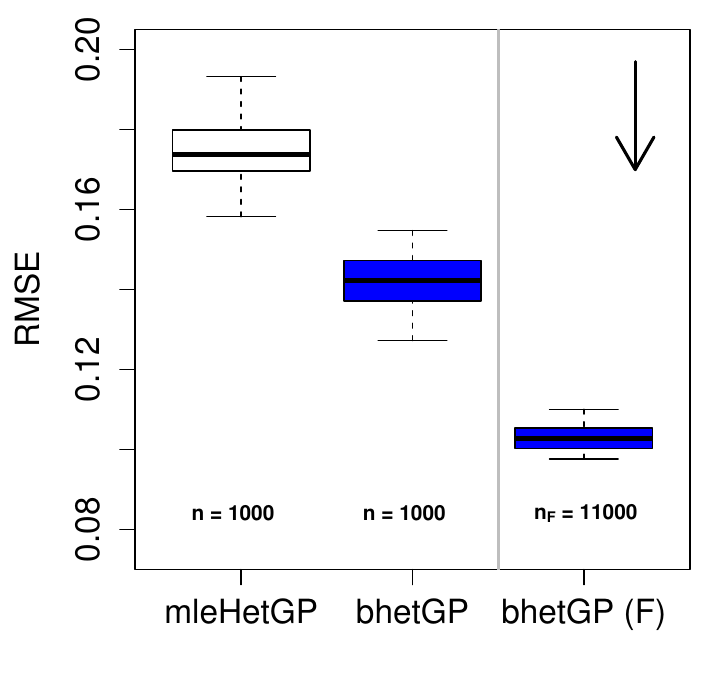}
    \includegraphics[scale = 0.5, trim=0 5 10 0, clip = TRUE]{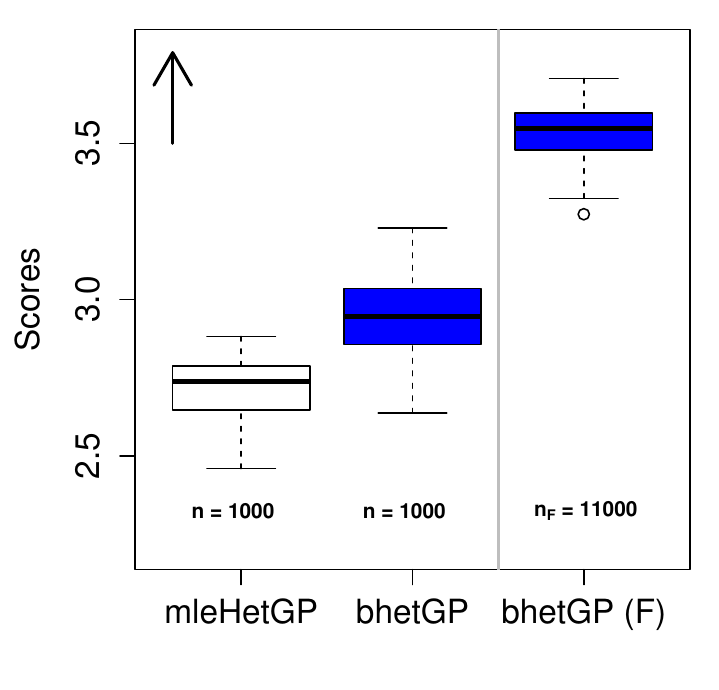}
    \caption{RMSE \textit{(left)} and score \textit{(right)} for summarized over
    fifty MC trials for {{\tt hetGP}, {\tt bhetGP}, and {\tt bhetGP (F)}
    using the larger training set with $n_F = 11,000$.} 
    Lower RMSEs and higher scores are preferred.}
    \label{fig:ato}
\end{figure}

Each of fifty MC trials proceed as follows: $n=1000$ training locations are
chosen at random from $n_{\mathrm{tot}}$, along with a random number of
replicates $a_i \sim \mathrm{Unif}(1, 2, \dots, 10)$, for $i=1,\dots,n$.  Then
$n_p = 1000$ testing locations are chosen from the complement, with all
replicates. {We fit {\tt hetGP} and {\tt bhetGP} to the training data
and evaluated the approaches} via RMSE and score on the testing set. Figure \ref{fig:ato}
shows the results.  See boxplots to left of the gray partitioning line in both
panels. Observe that the {\tt bhetGP} methods are more accurate and have
better UQ.

To the right of the gray partition in the panels is a second set of {\tt
bhetGP} fits with the ``full'' remainder of unused runs ($n_F =
n_{\mathrm{tot}} - n - n_p = 10{,}000$), with all replicates. An ordinary {\tt
hetGP} cannot be fit on data this big.  Observe that these {\tt bhetGP} fits
are able to use the extra training data to improve accuracy and UQ.

\subsection{NOAA-GLM lake temperature forecasts}
\label{sec:glm}

Forecasting problems present a major challenge in ecology and represent a
crucial component in studying and understanding ecological phenomena, both in
the near and long term \citep{clark2001ecological}. In particular, water
temperature forecasts \citep{wander2023data, thomas2023near} are a key factor
to crisis mitigation for resource management \citep{lee2023data,
radeloff2015rise}. We developed {\tt bhetGP} to support lake temperature
forecasting efforts. Fluctuations in temperature affect aquatic ecosystems in
many ways, and accurate predictions can be essential for management
\citep{woolway2021lake}. Higher temperatures promote {algal and
cyanobacterial growth}, which can
adversely affect drinking water \citep{carey2012eco, paerl2009climate}.

Process models comprise an important component in forecasting water quality
via chlorophyll level, temperature and their interactive dynamic
\citep{carey2022advancing}. Such models often require calibration
\citep{lofton2023progress} so that they capture the field data appropriately.
Our empirical exercise here focuses on forecasting lake temperatures at
Falling Creek Reservoir (FCR), located in Vinton, Virginia, USA. We have data
from a simulation campaign \citep{holthuijzen2024synthesizing} using the
General Lake Model \citep[GLM;][]{gmd-12-473-2019}, furnishing forecasts
of lake temperature along a water column at multiple depths, as driven by NOAA
weather forecasts. 

\begin{figure}[ht!] 
\centering
\includegraphics[scale=0.7, trim= 0 55 10 5, clip=TRUE]{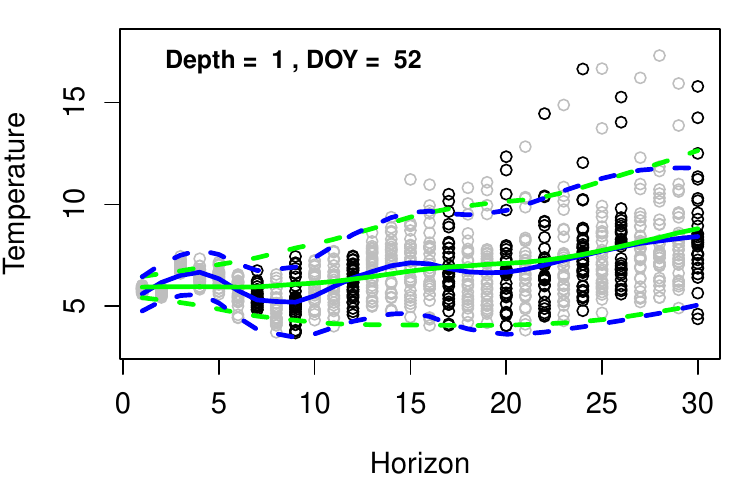}
\includegraphics[scale=0.7, trim= 25 55 10 5, clip=TRUE]{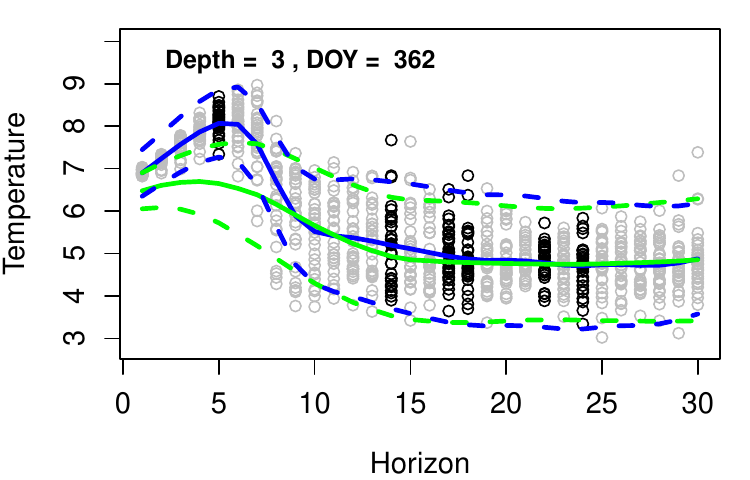}
\includegraphics[scale=0.7, trim= 0 1 10 5, clip=TRUE]{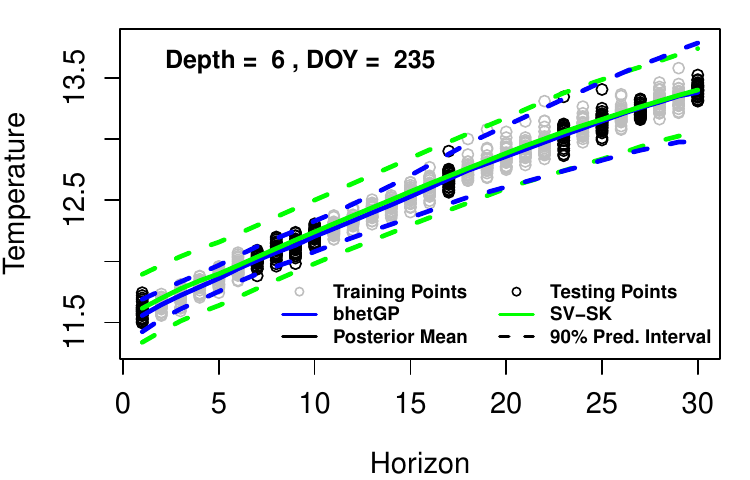}
\includegraphics[scale=0.7, trim= 25 1 10 5, clip=TRUE]{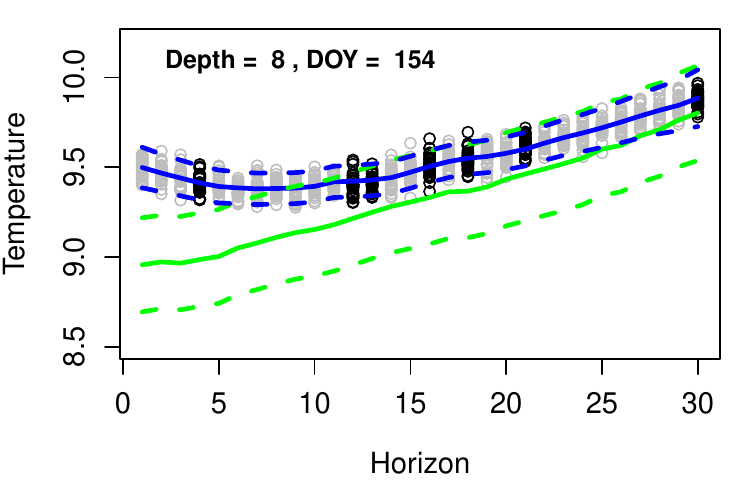}
\caption{Predictions across horizons for four days and depths.}
\label{fig:lakes_1d}
\end{figure}

Some example simulations, along with fits described momentarily, are shown in
Figure~\ref{fig:lakes_intro}.  Each panel plots temperature realizations (all
open circles, regardless of color) across a horizon of thirty days for a
particular day of year (DOY) and depth (in meters). GLM outputs are a
deterministic function of inputs, but when driven by a 31-member NOAA
weather ensemble it produces a spread of forecasts whose diversity generally
increases with horizon, especially at shallower depths. Our data set is
comprised of 968 days (3+ years) for ten depths and out thirty days into the
future.  So $n=968
\times 10 \times 30 = 290{,}400$ with 31-fold replication ($a_i = 31$
uniformly) so that $N\approx 9$ million.  Needless to say, ordinary {\tt
hetGP} is a non-starter.

\citeauthor{holthuijzen2024synthesizing}~collected these simulations and
developed an {economical} surrogate by combining scaled Vecchia fits using an
off-the-shelf library.  The idea is reminiscent of SK, coupling separate fits
to averages $\bar{y}_i$ and residual sums of squares $s_i$, similar to
Eq.~\eqref{eq:si}.  This provided a capability that was unmatched, and yielded
reasonably accurate forecasts for NOAA-GLM, but there were two downsides. One
is that the link between mean and variance was not explicit in the modeling.
Another is that the software learns by maximizing, so it does not provide a
full accounting of uncertainty.
\citeauthor{holthuijzen2024synthesizing}~worked around this, to a degree, but
taking an upper quartile of variance estimates -- an idea we appropriated with
$\Lambda(\mathcal{X}) =
\exp\{\mu_n^{\ell}(\mathcal{X}) + \Phi^{-1}_{0.95}
\sigma_n^{\ell}(\mathcal{X})\}$ {for this example}. This is important, as we shall illustrate,
but there is also much scope for improvement.

Eventually we intend to follow \citeauthor{holthuijzen2024synthesizing}~and
use {\tt bhetGP} in a bias-correcting and calibration context
\citep[e.g.,][]{kennedy2001bayesian}.  However,  our exercises here focus on
surrogate modeling and predictive capabilities only. Situating our surrogate
in a larger framework is part of future work, with other ideas in Section
\ref{sec:discussion}.  We use the $N\approx 9$M run campaign identically to
the original study, along with the Vecchia/SK-like code provided, but with one
exception.  Some temperature measurements at the lowest depths have
essentially no noise, which is both unrealistic and potentially problematic
when modeling variances with log transformations.  As a workaround we added
$\mathcal{N}(0, 0.01)$ jitter throughout.


Our exercise involves random 80:20 train:test splits of the unique-$n$ inputs,
taking all replicates.  In Figure \ref{fig:lakes_1d} the gray points are in
the training set and the black ones are held out for testing.  The figure
shows two fits, our default {\tt bhetGP} (blue) fit and
\citeauthor{holthuijzen2024synthesizing}'s default SK-like fit.
{Both use Vecchia approximations with $m = 50$ NNs based on
Euclidean distance on coded inputs. To streamline our graphics, we do not
explicitly notate the use of Vecchia.} Notice how both methods are mostly in
agreement with the exception of the {\em bottom-right} panel.  Yet {\tt
bhetGP} is both more accurate and has a narrower predictive interval (PI).
Occasionally, the SK-like method has a ``bad day'', where it's substantially
off in both mean and uncertainty. {It worth noting that
increasing $m$ can lead to better predictions, at the expense of computational
resources.}

\begin{figure}[ht!]
\centering
\includegraphics[height=0.3\textwidth, trim= 2 4 10 0, clip=TRUE]{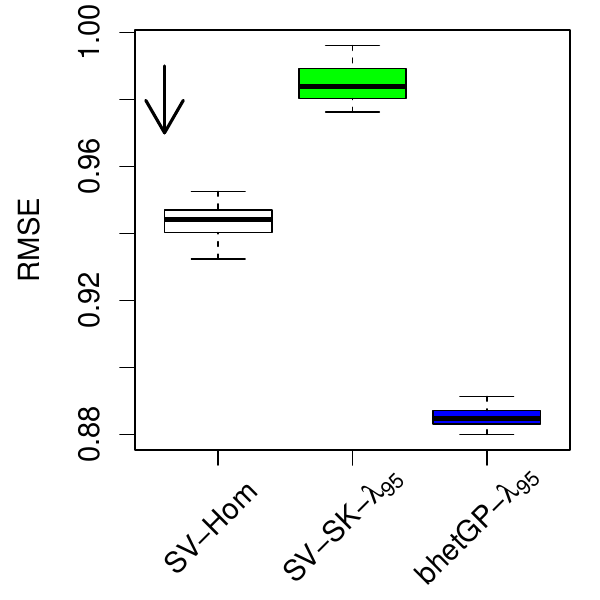}
\includegraphics[height=0.3\textwidth, trim= 2 4 10 0, clip=TRUE]{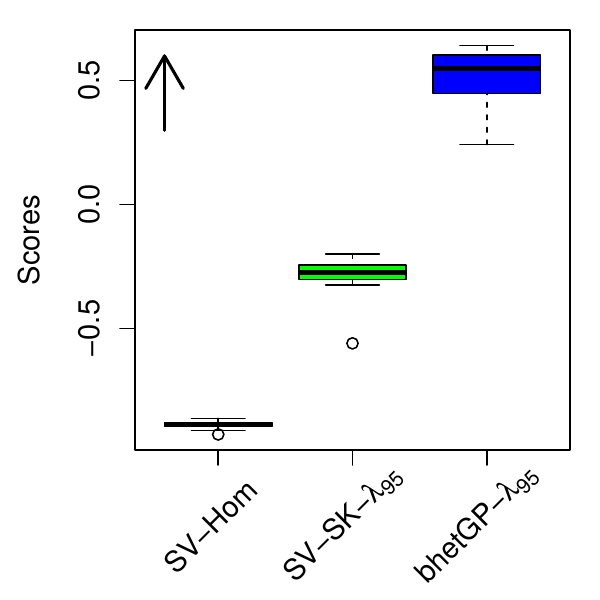}
\caption{RMSE ({\em left}, lower is better) and scores ({\em
right}, higher is better) over  thirty MC repetitions.}
\label{fig:lakes_results}
\end{figure}

\begin{figure}[ht!] 
\centering
\includegraphics[scale=0.62, trim= 0 0 7 5, clip=TRUE]{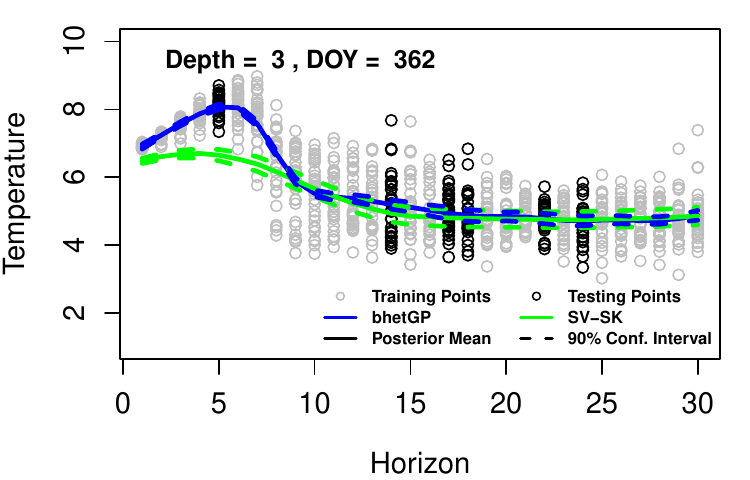}
\includegraphics[scale=0.38, trim= 35 0 7 5, clip=TRUE]{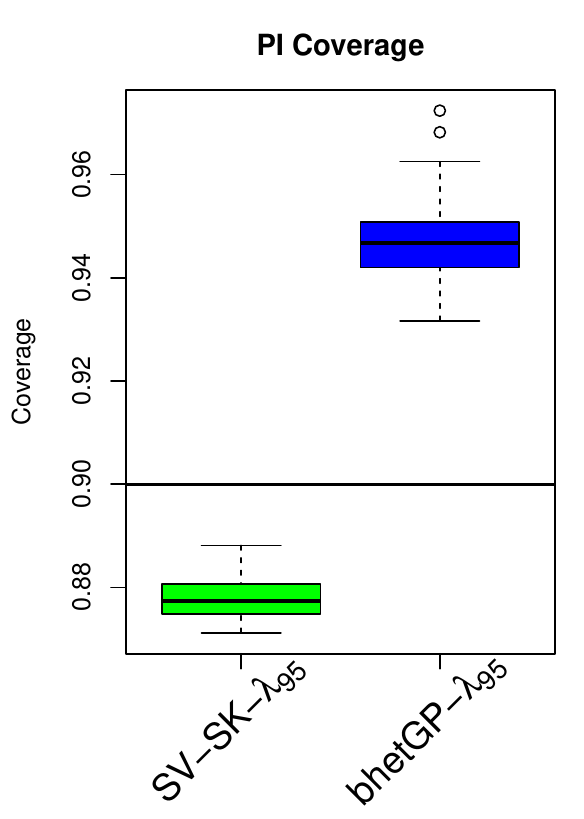}
\includegraphics[scale=0.38, trim= 35 0 7 5, clip=TRUE]{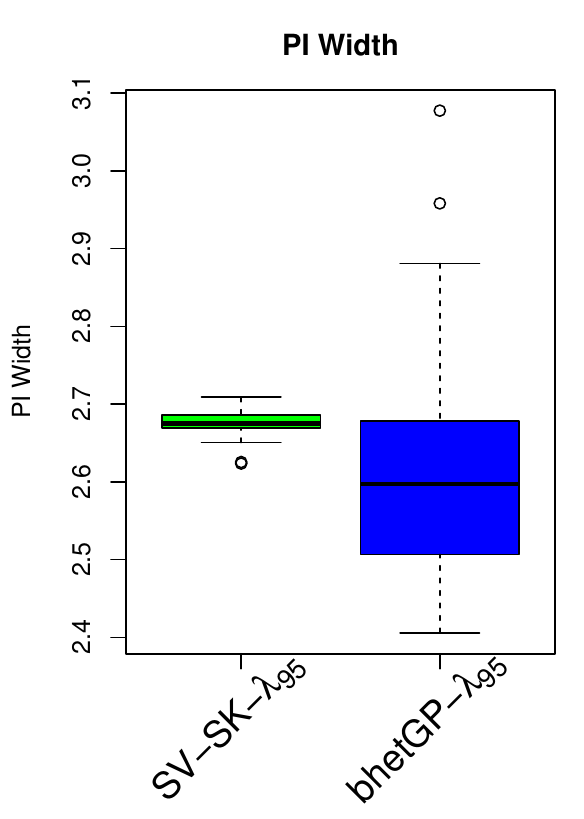}
\includegraphics[scale=0.38, trim= 35 0 7 5, clip=TRUE]{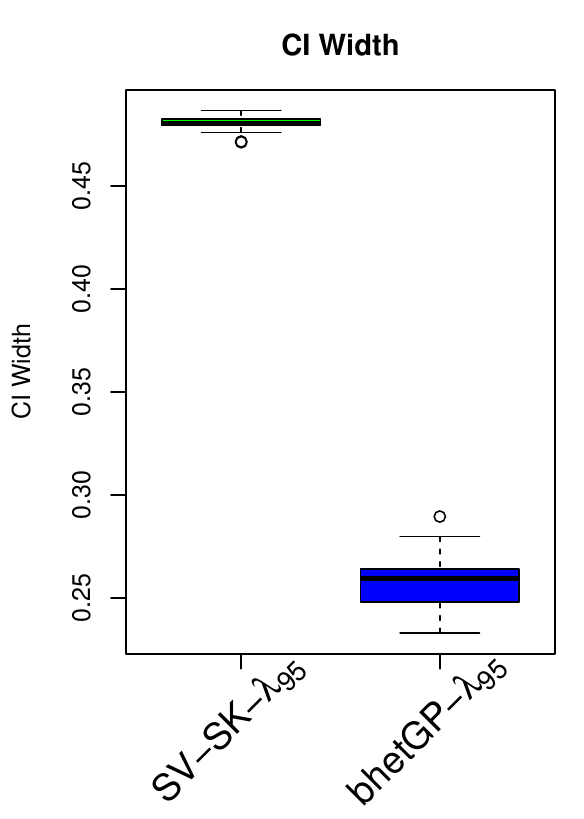}
\caption{{\em Left to right:} Mean predictions with CI for a particular day
(362) and depth (3m), coverage based on 90\% PIs along with mean widths of PI
and CI.} 
\label{fig:lakes_ci}
\end{figure}

Figure \ref{fig:lakes_results} shows out-of-sample RMSE and score results from
thirty MC repetitions of random train:test splits.  For perspective here, we
have added an ordinary {(homoskedastic/non-SK)} scaled Vecchia fit using
software defaults.  Observe that both heteroskedastic fits dramatically
outperform that baseline in terms of scores.  
Our {\tt bhetGP} consistently outperforms the
previous SK-like method.  Finally, Figure \ref{fig:lakes_ci} takes a closer
look at UQ via 90\% intervals: both predictive (PI) summarizing full
uncertainty, and confidence (CI) focusing just on mean uncertainty.  CIs
comprised an important aspect of the
\citeauthor{holthuijzen2024synthesizing}~analysis, as it was only the mean
prediction that was used downstream in their calibration.  To obtain CI in
{\tt bhetGP}, use Eq.~\eqref{eq:predy} and drop the $\lambda(\mathcal{X})$
term. First focus on the {\em left} panel, which shows the CI corresponding to
the {\em top-right} panel of Figure \ref{fig:lakes_1d}. (The means shown are
the same.)  Notice that {\tt bhetGP}'s interval is much narrower 
{and fits the data better}, which is
consistent across other days/depths and MC repetitions, as the right-most
panels show for both CIs and PIs.

{In the second ``PI coverage'' panel, the suffix requires additional
explanation}.  CI coverage can't
be evaluated here because we don't know what the true means are. However, the
full set of held-out replicates can be used to assess the quality of PIs.  We
simply tabulate the proportion which lie inside the interval and compare that
to the nominal rate (90\%).  {For this example, we use 
$\Lambda(\mathcal{X}) = \exp\{\mu_n^{\ell}(\mathcal{X}) + \Phi^{-1}_{0.95}
\sigma_n^{\ell}(\mathcal{X})\}$, i.e., the upper quantile for noise levels,
to obtain bounds.  Appendix \ref{app: glm} provides evidence that using mean
noise levels undercuts uncertainty. Observe that, with a conservative noise
level, {\tt bhetGP} provides a much higher coverage rate whereas SV-SK falls
short despite having wider interval widths on average across the runs. }


\section{Discussion}  \label{sec:discussion}

We proposed a Bayesian heteroskedastic Gaussian processes ({\tt bhetGP}) model
synthesizing \citet{binois2018practical}'s Woodbury likelihood (ordinary {\tt
hetGP}) and \citet{sauer2023vecchia}'s Vecchia approximated latent inference
via elliptical slice sampling (ESS).  The result is a new capability that can
furnish full posterior inference/prediction in a workable implementation,
computationally speaking.  Our {\tt bhetGP} was demonstrated to be faster and
more accurate than an ordinary {\tt hetGP}.  It also beats a recently proposed
stochastic krigning (SK) analog proposed \citep{holthuijzen2024synthesizing}
to model a massive {campaign} of lake temperature simulations.

We envision several potential extensions.  For example, many stochastic
simulation settings involve noise levels that change only for subset
of inputs (being constant across the others).  It may be advantageous to
allow a user to specify on which coordinates the noise varies.
{Although our method infers longer lengthscales for dimensions
that do not largely contribute to noise levels, specifying such dynamics {\em
a priori} can save computational time and prevent overfitting of the noise
process.} Another possible extension involves further hybridization with
\citet{sauer2023vecchia}'s deep GP framework, coupling input warping with
input-dependent noise. { We also see scope for tuning $m$ in
the Vecchia approximation, rather than using heuristic values, to maximize
predictive accuracy and fidelty under a fixed computational budget.}

Finally, surrogate models aren't just predictors but usually a means to an end
for some downstream task, like computer model calibration
\citep{kennedy2001bayesian}, active learning \citep{binois2018replication} or
Bayesian optimization \citep{jones1998efficient}.  A key component in each of
those examples is that UQ plays an outsized role.  We plan to investigate {\tt
bhetGP}'s potential for calibrating NOAA-GLM simulations to sensor
observations \citep{holthuijzen2024synthesizing}, and remain optimistic about
{its} potential in many other settings.

\subsection*{Acknowledgments}

We are grateful to the NSF for funding (\#2318861), and members of the Rules of
Life project, specifically Mary E.~Lofton for NOAA-GLM lake temperature
forecasting data products. We also thank Maike Holthuijzen for the lake
temperature data and other support.

\bibliographystyle{jasa}
\bibliography{references}

\newpage
\begin{center}
{\large\bf Appendix}
\end{center}
\appendix

\section{MCMC via MH v/s ESS}\label{app: MH}

{Figure~\ref{fig:mh-ess} provides a comparison between MCMC
sampling via MH and ESS for the 1D motorcycle example, both initialized at
the same values (black line), and with fixed MLE estimates for
$\theta_\lambda$ and $\theta_y$.
\begin{figure}[ht!]
    \centering
    \includegraphics[scale = 0.5, trim=2 0 0 0, clip = TRUE]{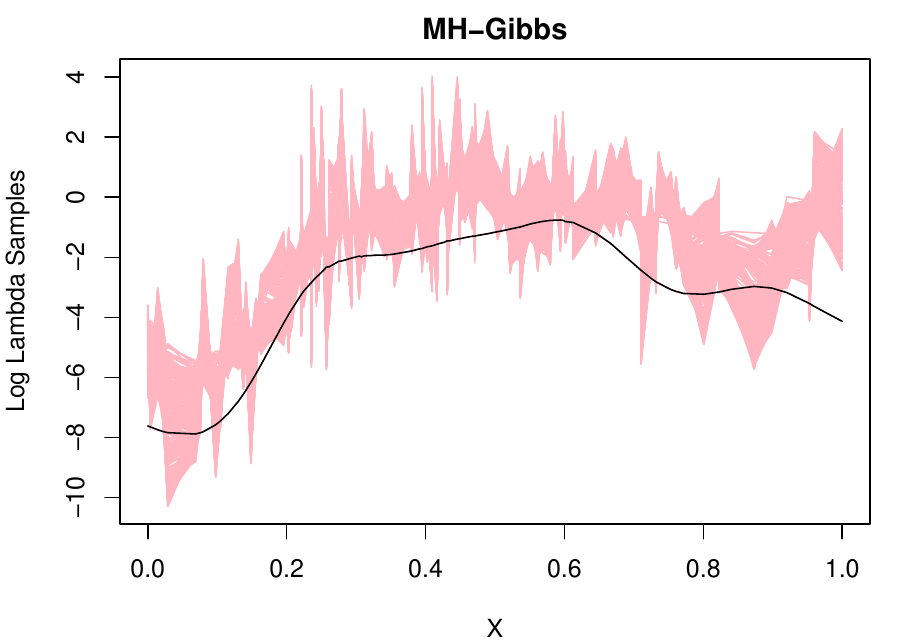}
   \includegraphics[scale = 0.5, trim=15 0 0 0, clip = TRUE]{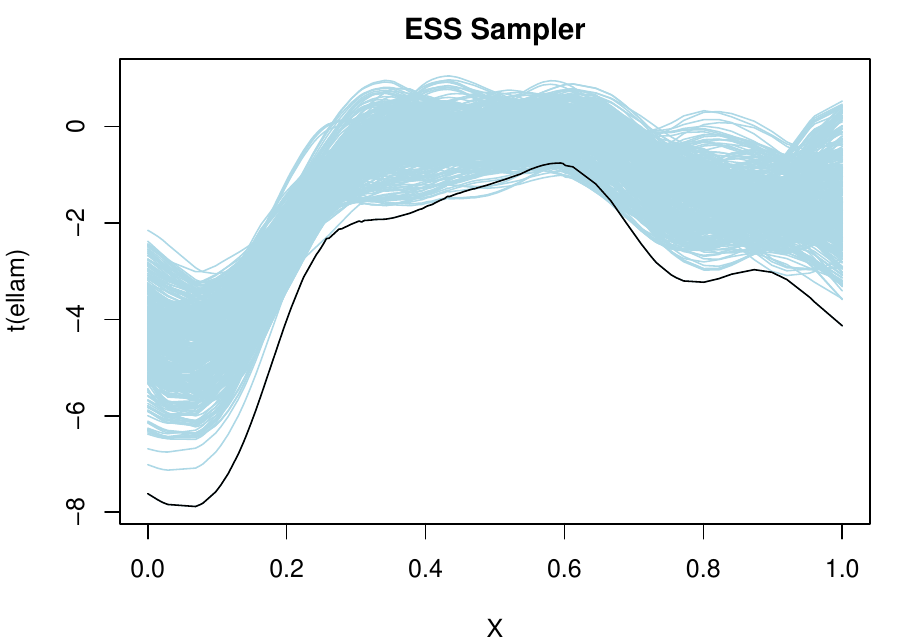}
    \includegraphics[scale = 0.5, trim=2 0 0 0, clip = TRUE]{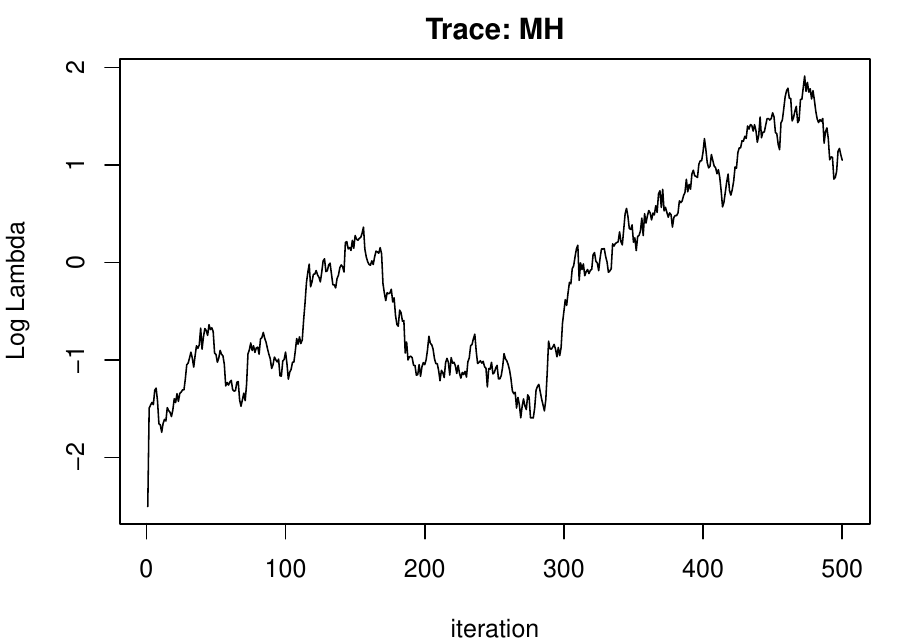}
   \includegraphics[scale = 0.5, trim=15 0 0 0, clip = TRUE]{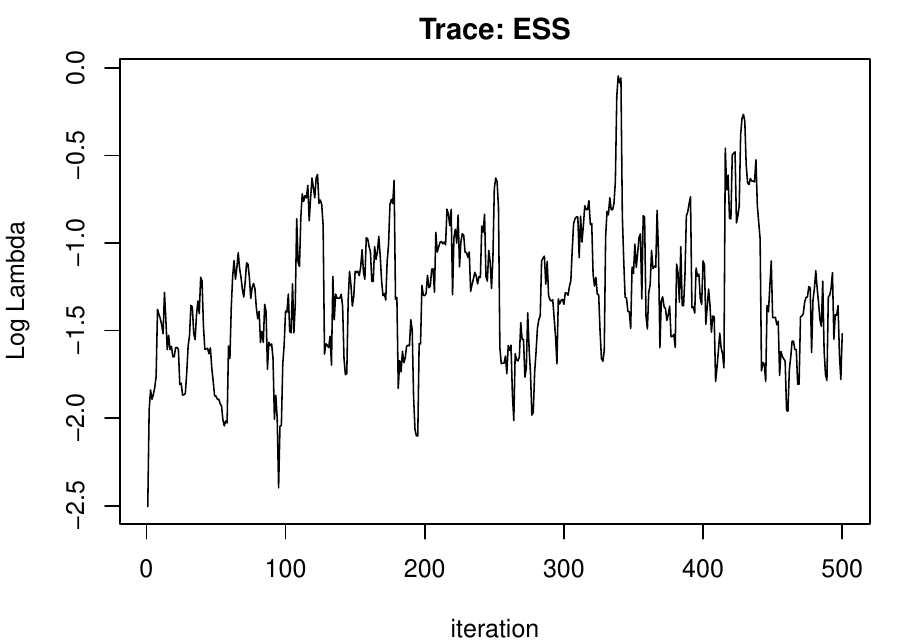}
    \caption{ {(\it top)} $\log \Lambda$ samples obtained via MH and ESS for 1D motorcycle problem. {(\it bottom)} Trace plot for one
    $\log \Lambda_i$ to evaluate convergence.}
    \label{fig:mh-ess}
\end{figure}
Observe in the top panels that the noise process is not smooth via MH due to
independent sampling of each $\lambda_i$.  This is inconsistent with a lemma
from \cite{binois2018practical} which states that the likelihood is maximized
when $g_\lambda = 0$. In contrast, ESS produced smooth samples. Additionally
the MH trace plot, shown in the bottom panel(s) for a particular $\lambda_i$,
indicates that the MCMC has not fully converged.  The trace plot for ESS
sampling is much better behaved. Run times for MH and ESS were 262.8 seconds
compared to 7.1 seconds, respectively. For a small enough problem, as is the
case with the motorcycle example, even the slower of the two is manageable.
However, as $n$ increases, and dimensionality $d$ of input space increases, MH
for latent $\lambda$-values is not a viable alternative.}

\section{SVSHGP v/s bhetGP}\label{app:py}

\begin{figure}[ht!]
    \centering
    \includegraphics[scale = 0.6, trim=2 0 5 10, clip = TRUE]{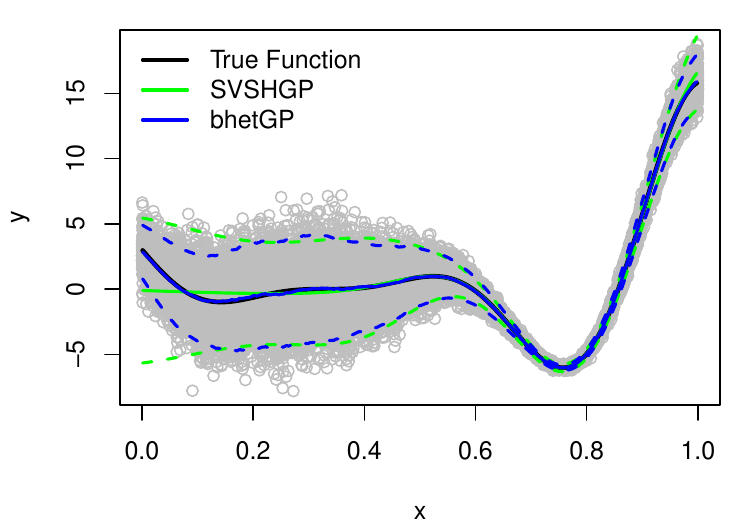}
    \includegraphics[scale = 0.6, trim=2 0 5 10, clip = TRUE]{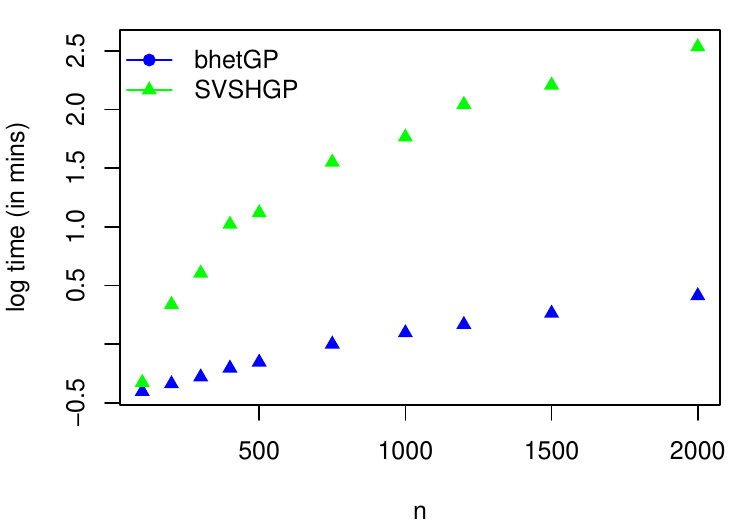}
    \includegraphics[scale = 0.6, trim=2 0 5 10, clip = TRUE]{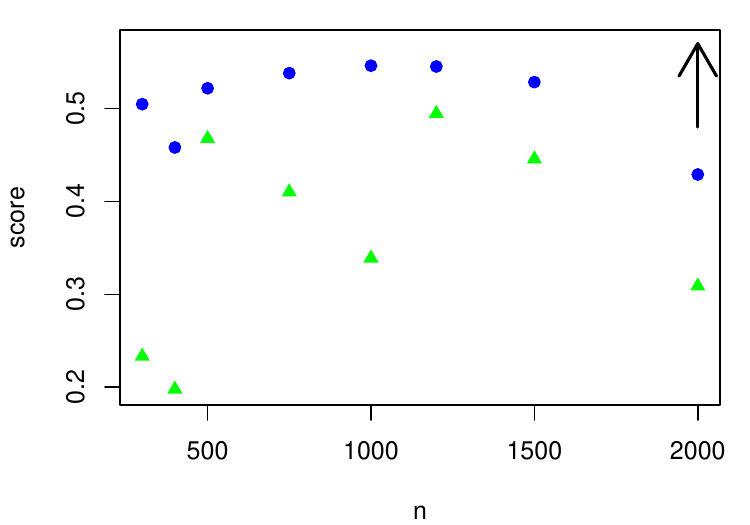}
    \includegraphics[scale = 0.6, trim=2 0 5 10, clip = TRUE]{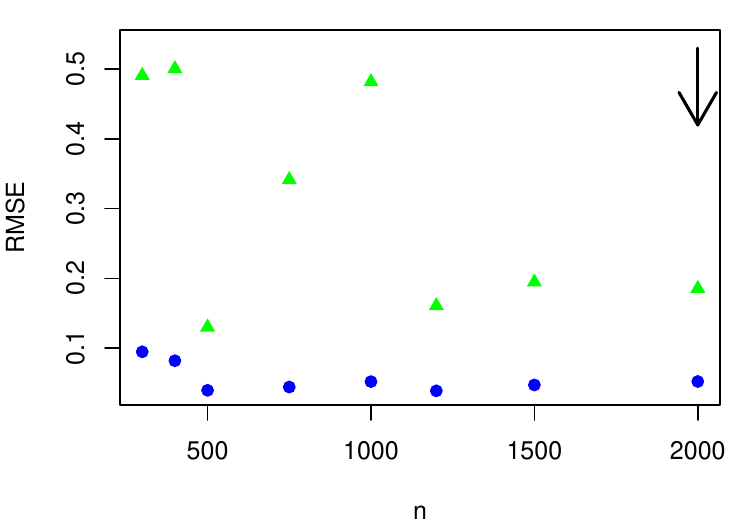}
    \caption{ {\em(top)} {\em Left:} {\tt bhetGP} and SVSHGP for $n = 1000$ and $N=30n$. 
    {\em Right:} time in log minutes varying $n$. {\em(bottom)} {\em Left:}
     Scores and {\em Right:} RMSEs for SVSHGP and {\tt bhetGP}.}
    \label{fig:pycom}
\end{figure}

{ We compare {\tt bhetGP} to SVSHGP \citep{liu2020large} from
the {\sf Python} library {\tt GPflow}. SVSHGP leverages inducing points (IPs)
for large-scale approximations and combines two GPs \citep{saul2016chained}
for mean and noise process estimation. To contrast their performance, we use
the 1D toy example from Section \ref{sec:vecn} with a variety of $n$-values
and $N=30n$.  Throughout we use $m=10$ inducing points or Vecchia NNs,
respectively.  See the panels of Figure~\ref{fig:pycom}.}

{Begin with the {\em top-left} panel, which shows two example
fits. Observe that {\tt bhetGP} captures the true function almost perfectly
and provides good UQ across the range of $x$. SVSHGP provides similar
predictions except in the left half of the input space where it over-smooths.
The UQ is off for $x < 0.1$.  The remaining panels, clock-wise from the {\em
top-right} illustrate how {\tt bhetGP} outperforms SVSHGP over all $n$ in
terms of time, RMSE and score, respectively.}

{ To help explain why {\tt bhetGP} is able to best SVSHGP it's
worth remarking on a connection between inducing points and our unique-$n$
representation. When $m = n$ inducing points are used, and when they are
placed at exactly the unique-$n$ locations, the two methods are identical.
Under the hood, the exact same Woodbury trick is used to speed-up computation,
and both provide exact calculations for the likelihood and prediction
\citep{binois2018practical}.  However, when $n$ (and thus $m=n$) is too big to
work with, computationally speaking, Vecchia offers a better approximation
framework.  This is true not only because one need not choose (possibly
poorly) the location of those $m$ inducing points \citep{garton2020knot}, but
because the Vecchia approximation is more localized.  Compared to Vecchia,
inducing point approximations tend to be ``blurry'', i.e., low-fidelity, which
has been remarked on at length in recent literature
\citep[e.g.,][]{wu2022variational,sauer2023vecchia}.}

\section{NOAA-GLM lake temperature forecasts: Coverage}\label{app: glm}

\begin{figure}[ht!]
    \centering
    \includegraphics[scale = 0.445, trim=0 80 14 10, clip = TRUE]{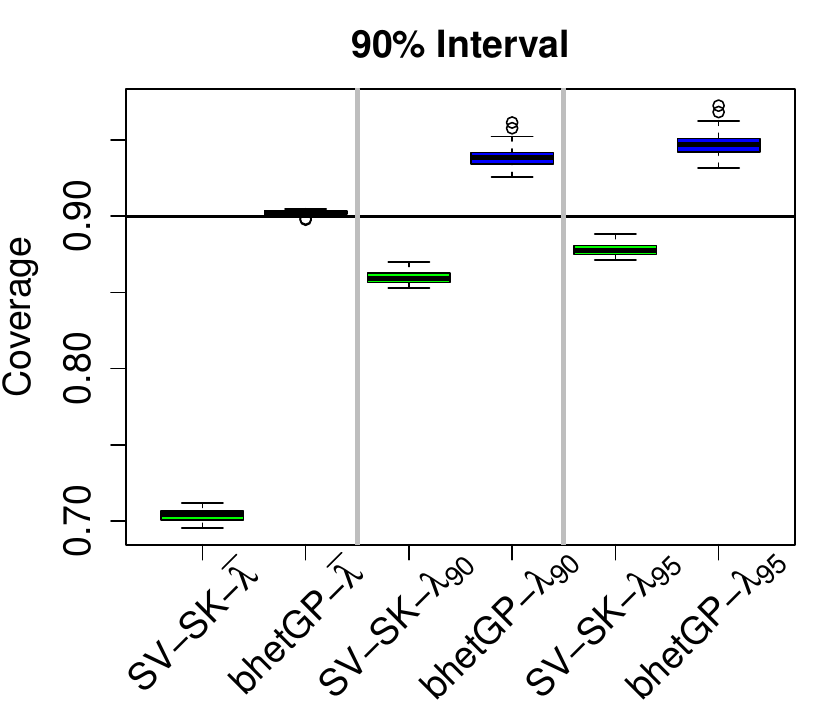}
   \includegraphics[scale = 0.445, trim=20 80 14 10, clip = TRUE]{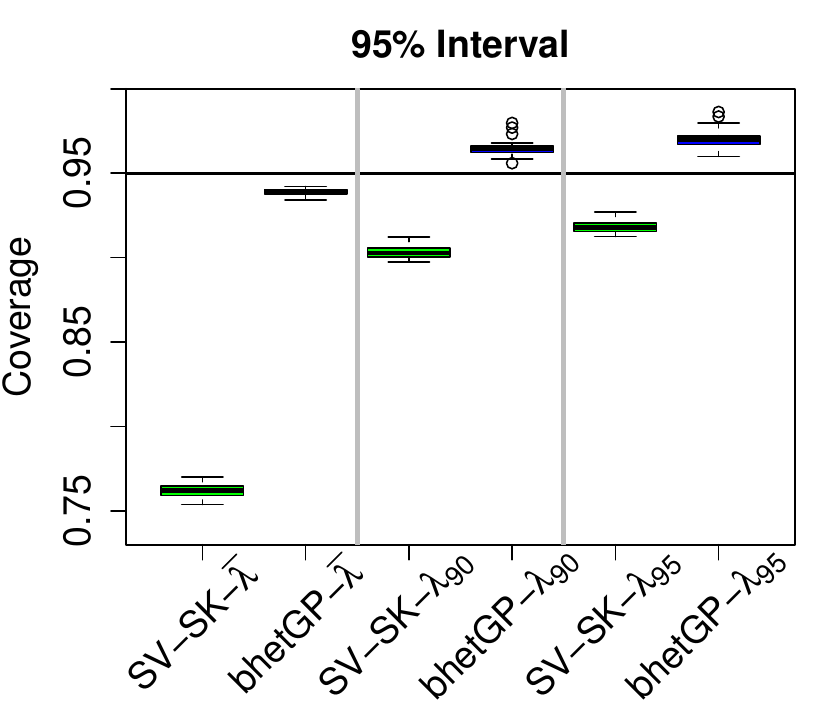}
    \includegraphics[scale = 0.445, trim=20 80 14 10, clip = TRUE]{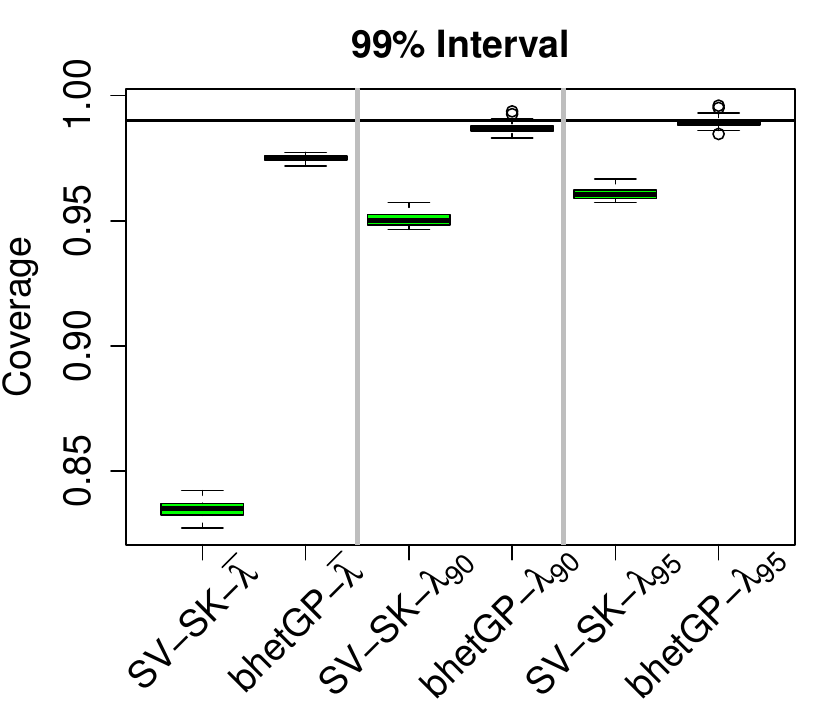}
    \includegraphics[scale = 0.445, trim=0 0 14 30, clip = TRUE]{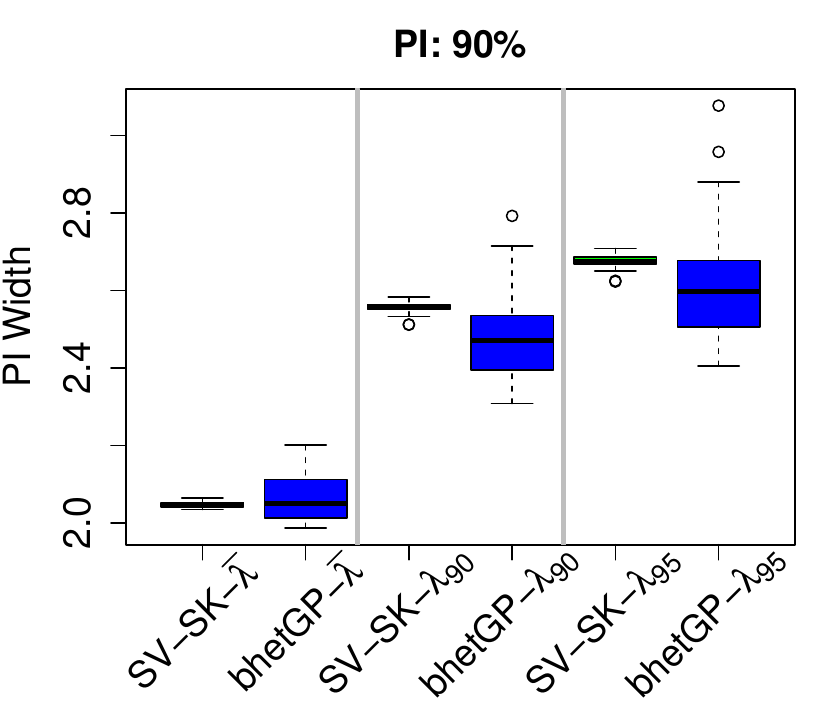}
    \includegraphics[scale = 0.445, trim=20 0 14 30, clip = TRUE]{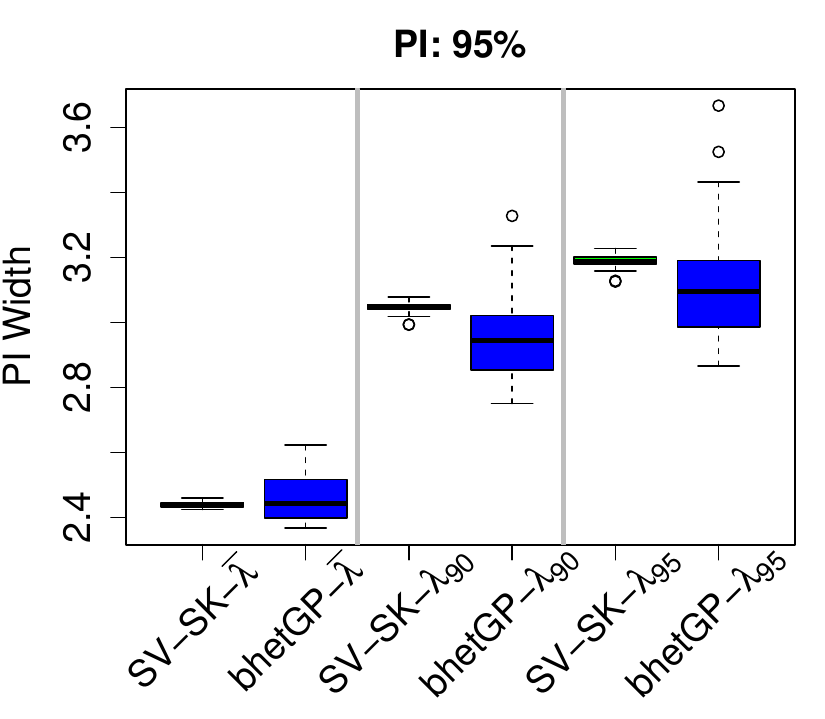}
    \includegraphics[scale = 0.445, trim=20 0 14 30, clip = TRUE]{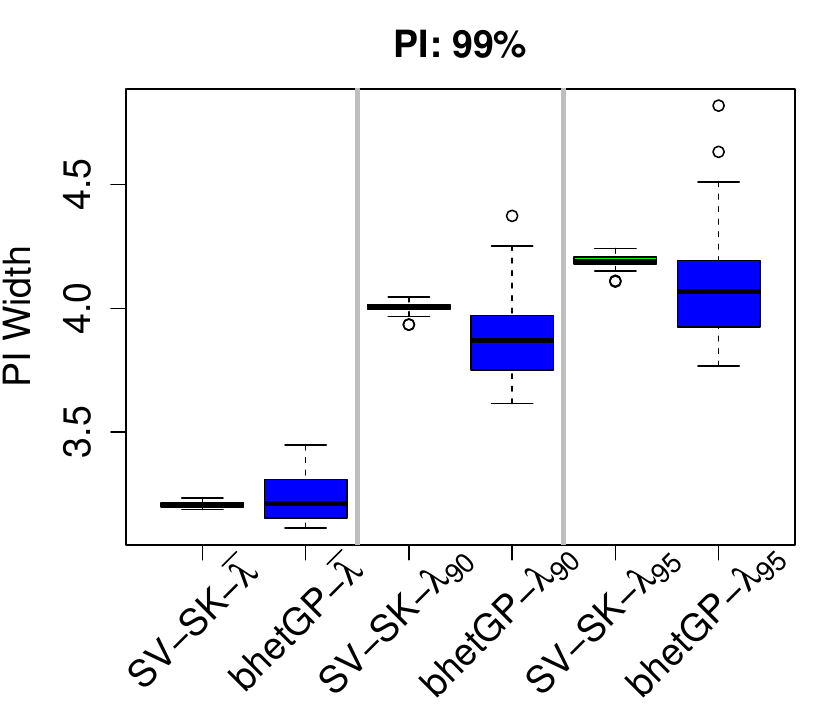}
    \caption{ {(\it top)} Coverage and {(\it bottom)} width of intervals for levels 90\%, 95\% and 99\% using 0.5, 0.9 and 0.95 quantiles of noise levels to calculate the bounds for SV-SK and bhetGP.}
    \label{fig:lakescov}
\end{figure}
{Here we provide a comparison of coverage for nominal levels
$90\%, 95\%$ and $99\%$, as predicted by three estimated levels: mean
($\bar{\lambda} \equiv 50\%$ quantile), upper 90\% quantile for the latent
noise process ($\lambda_{90}$), and upper 95\% quantile ($\lambda_{95}$). See
Figure~\ref{fig:lakescov}. Observe that {\tt bhetGP}'s {(\it blue)}
empirical coverage is always close to nominal.  Occasionally, $\bar{\lambda}$
falls short at 95\% and 99\% intervals. SV-SK {(\it green)} always under
covers, with $\bar{\lambda}$ providing a very low coverage across all interval
levels. Notice that in terms of widths, {\tt bhetGP} has larger variation
across MC repetitions, but is on average narrower.}

\begin{figure}[ht!]
    \centering
    \includegraphics[scale = 0.54, trim=0 20 4 0, clip = TRUE]{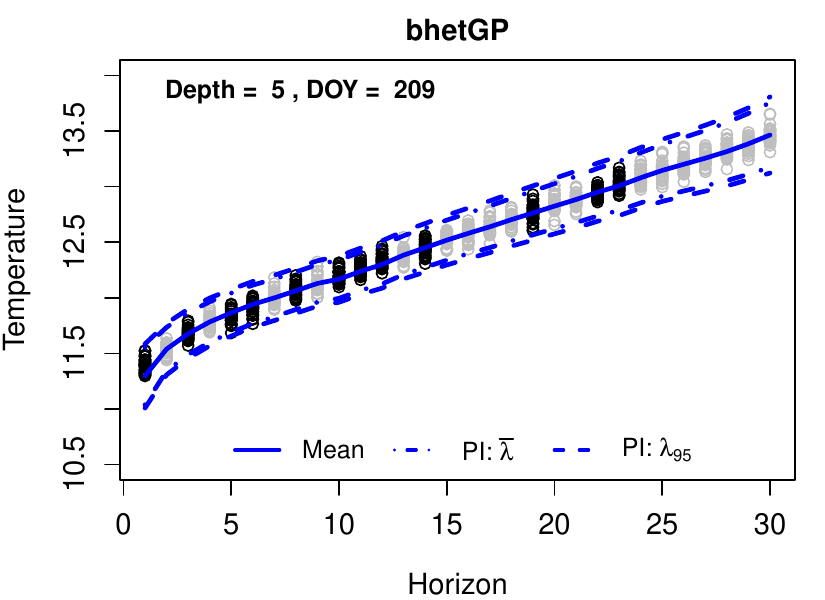}
   \includegraphics[scale = 0.54, trim=18 20 4 0, clip = TRUE]{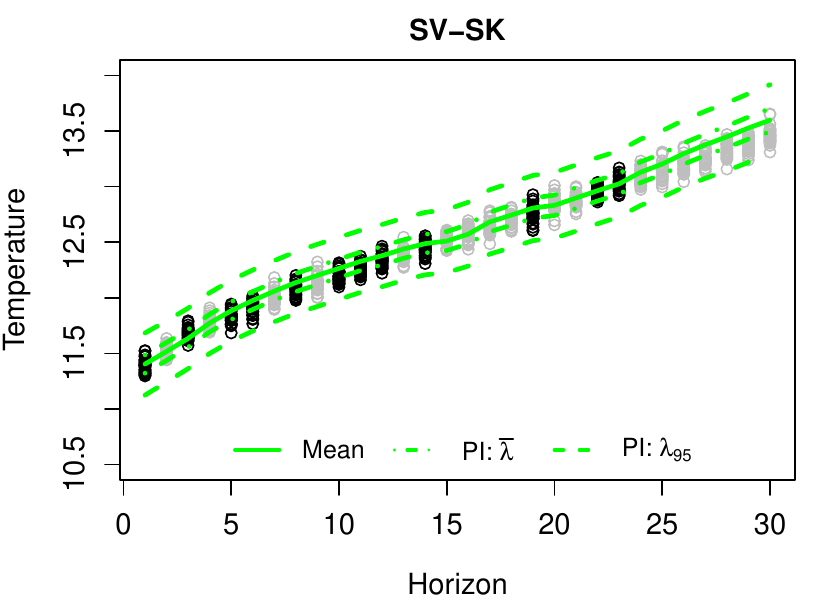}
    \includegraphics[scale = 0.54, trim=0 0 8 20, clip = TRUE]{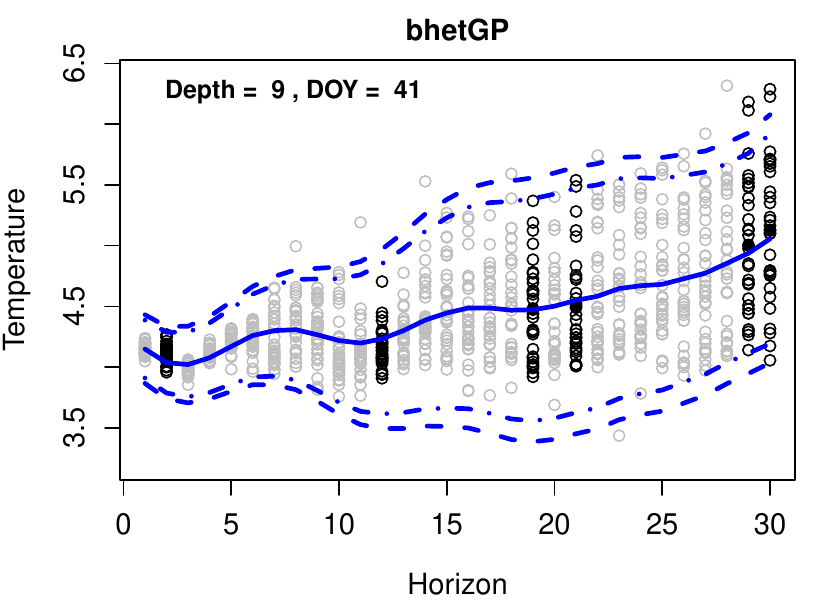}
    \includegraphics[scale = 0.54, trim=18 0 8 20, clip = TRUE]{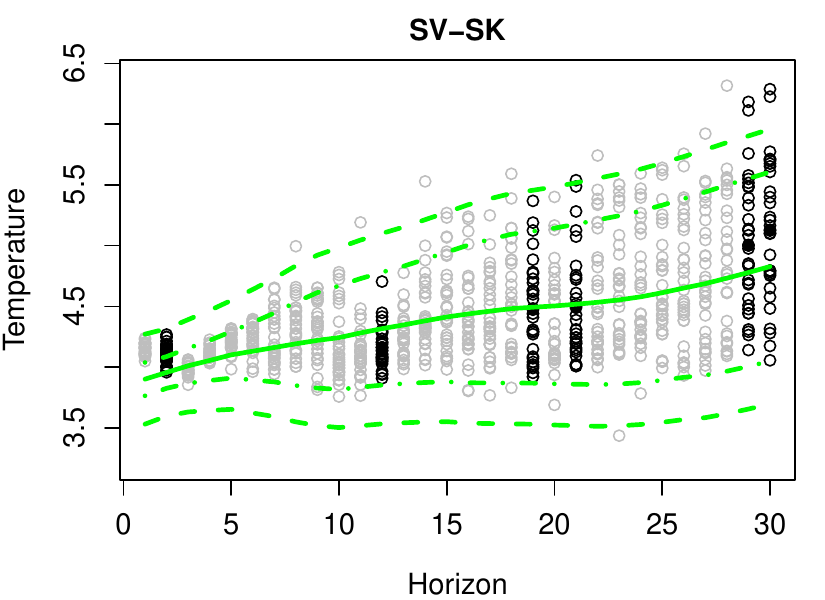}
    \caption{Predictions for 2 days of year using bounds calculated with 0.5
    and 0.95 quantile estimates for $\log \lambda$ for bhetGP {(\it left)} and
    SK-SV {(\it bottom)} }
    \label{fig:lakes50v95}
\end{figure}

{Now consider a comparison between {\tt bhetGP} and SV-SK 90\%
prediction intervals using two different noise estimates: $\bar{\lambda}$
and $\lambda_{95}$.  See Figure~\ref{fig:lakes50v95}.}
{Our {\tt bhetGP} does not substantially benefit from using a
conservative noise estimate. However, SV-SK has particularly poor coverage
without via $\bar{\lambda}$, leading to a PI that misses about half of the
out-of-sample responses.}

\end{document}